\documentclass[trackchanges,twocolumn]{aastex702}

\usepackage{amsmath}
\usepackage{subcaption}

\usepackage[normalem]{ulem}

\begin{document}

\title{Analyzing Fermionic Dark Matter Scenarios with Anomalous Compact Objects}

\correspondingauthor{Yaiza Cano}
\author[0009-0004-3312-6694,gname=Yaiza,sname=Cano]{Yaiza Cano}
\affiliation{Universidad de Alcal\'a, Grupo de F\'isica Nuclear,  Part\'iculas y Astrof\'isica, \\ Departamento de F\'isica y
Matem\'aticas, 28805 Alcal\'a de Henares (Madrid), Spain}

\email[show]{ycano01@ucm.es}  

\author[0000-0002-7764-9823,gname=Jose M., sname=Alarcón]{Jose Manuel Alarc\'on} 

\affiliation{Universidad de Alcal\'a, Grupo de F\'isica Nuclear,  Part\'iculas y Astrof\'isica, \\ Departamento de F\'isica y
Matem\'aticas, 28805 Alcal\'a de Henares (Madrid), Spain}
\email{jmanuel.alarcon@uah.es}

\begin{abstract}

In this paper, we consider three compact objects (HESS J1731-347,  PSR J1231-1411, XTE J1814-338) with anomalous mass-radius relations to analyze the possibility of their being dark matter admixed neutron stars. 
We try to infer the properties of the dark matter particles, under the assumption of behaving as a free Fermi gas. The main novelty relies on the use of a baryonic equation of state obtained from first principles over the whole density range, which allows the model dependence of the baryonic part of the calculation to be eliminated. Once the possible dark matter admixed neutron star configurations are obtained, we check their stability and whether it is feasible for a neutron star to capture the necessary dark matter fraction. 
Our analysis shows that our dark matter model favours the dark matter admixed neutron star hypothesis over the pure baryonic neutron star one, with the most favoured dark matter mass of $m_{DM} = 1$~GeV, although the necessary dark matter content cannot be explained in a standard galactic environment.
\end{abstract}


\section{\textbf{Introduction}}

\color{black}
Detection of dark matter (DM) is one of the most important tasks in modern physics. 
The large-scale structure of the Universe and its evolution are believed to be greatly influenced by the gravitational effects produced by this type of matter. By now, there is plenty of cosmological evidence for its existence, but because of the weak interactions between baryonic and dark matter particles, searches for both indirect and direct detection have so far been fruitless. 
This makes constraining the parameter space of DM one of the most important tasks of modern physics.
On the other hand, DM might be present in compact objects through capture mechanisms or mixed with the baryonic matter during the formation of the object. 
Neutron stars (NSs) are among the most compact objects in our Universe, after black holes, and are followed by white dwarfs. Because of their extreme densities, NSs provide
conditions to explore new phenomena, making them laboratories of new physics. In particular, the presence of DM inside an NS modifies its properties, providing an opportunity to study different DM scenarios.

Dark matter in NSs can manifest as a halo extending beyond the baryonic radius, which increases both the gravitational mass and the tidal deformability, or as a dark matter core inside the star, which decreases both quantities \citep{Nelson:2018xtr,Sagun:2021oml,Diedrichs:2023trk,Jockel:2023rrm}. 
In either case, having DM bound to an NS would offer different observables that could provide insight into the DM component of the star. Among the possible observational signatures are shifts in the merger frequency of gravitational waves, modifications to the X-ray pulse profile, and the impact of self-annihilating DM on the effective temperature, luminosity, and cooling history of the star \cite{Kouvaris:2007ay,Bhat:2019tnz,Cermeno:2017ejm}. Additional potential probes include changes in the kinematic behavior and rotational properties of the neutron star \cite{Perez-Garcia:2011tqq}. Moreover, DM could theoretically accumulate enough matter in an NS to collapse into a black hole, allowing black hole observations to constrain dark matter properties \cite{Bertone:2007ae, Kouvaris:2011fi, Kouvaris:2010jy}.

For this strategy to be successful, it is necessary to have reliable results for the properties of the compact object due to its baryonic content.
If the observables used are the mass and radius of the NS, the relevant theoretical input is the equation of state (EoS) of the baryonic matter.
Searches for the NS EoS have also been a major mission for physicists for the past decades, and astrophysical measurements could  tightly constrain the allowed EoS.
In this regard, there are new approaches to calculate the EoS of nuclear matter based on first principles \cite{Alarcon:2024hlj}. 
This reduces the model dependence in the baryonic part of the calculation, and the systematic error in the determination of the DM particle properties. 

This progress has immediate applications in searches for new physics where the baryonic EoS plays a central role. 
It allows one to rigorously quantify the effect of DM on the properties of the NS, and therefore constrain DM models and properties. 
In this paper we consider three compact objects located in the Milky Way: HESS J1731-347 ($R=10.4_{-0.78}^{+0.86}$~km,  $M=0.77_{-0.17}^{+0.20}$~M$_\odot$) \cite{2022NatAs...6.1444D}, PSR J1231-1411 ($R=9.91_{-0.86}^{+0.86}$~km, $M=1.12_{-0.07}^{+0.07}$~M$_\odot$) \cite{Qi_2025}
and XTE J1814-338 ($R=7.0_{-0.4}^{+0.4}$~km, $M=1.21_{-0.05}^{+0.05}$~M$_\odot$)\cite{Kini:2024ggu}. All these objects show masses and radii that deviate from the mass-radius relation predicted by the standard baryonic EoS and have been studied under strange-quark-star \cite{DiClemente:2024jph,Hong:2024qsc} or hybrid-star hypotheses \cite{Laskos-Patkos:2024fdp,Mariani:2024hybrid,Li:2025twin,Brodie:2023pjw}.
These anomalies make them candidates for being baryonic objects mixed with some DM content. 
This analysis also allows us to constrain the DM parameter space.

The effects of DM on compact objects (COs) have been studied for multiple combinations of dark and baryonic equations of state. 
For the dark matter EoS, there are studies with fermionic dark matter models (free Fermi gas \cite{Liu:2024swd,Kain:2021hpk}; self-interacting DM \cite{Barbat:2024yvi,Miao:2022rqj}; and interacting DM \cite{Issifu:2025qqw}) and bosonic dark matter \cite{Liu:2024swd,Tangphati:2025phx,Santos:2025xep,Shakeri:2022dwg,Shahrbaf:2024gdm,Shahrbaf:2025hsw,Karkevandi:2024vov,Karkevandi:2021ygv}. 
Some nuclear EoSs used in previous works are nonrelativistic phenomenological mean-field models \cite{Liu:2025cwy} (BSk22) and relativistic mean-field models \cite{Issifu:2025qqw} (DDME2 parameterization) and \cite{Lam:2025for} (DD2 parameterization). 
In addition, chiral effective theory has been used in EoSs describing NSs in the context of DANSs \cite{Barbat:2024yvi,Dengler:2025ntz,Deliyergiyev:2019vti,Dengler:2021qcq,Rutherford:2022xeb,Rutherford:2024uix}.

The paper is organized as follows. In Sec.~\ref{Sec:Methodology} we explain the EoSs used for baryonic and dark matter. In Sec.~\ref{Sec:Results} we show the results obtained for the different scenarios of DM content in the NS for different DM particle masses. Later, in Sec.~\ref{Sec:DM_content} we analyze the feasibility of the scenarios that could explain the anomalous CO masses and radii. Finally, in Sec.~\ref{Sec:Summary} we summarize the results and present the conclusions of our analysis.

\section{\textbf{Methodology}}\label{Sec:Methodology}
\subsection{Baryonic Matter EoS}

For the baryonic EoS we use the results obtained in Ref.~\cite{Alarcon:2024hlj}. 
This calculation uses regulator-independent results  of effective fiel theores for the EoS at low densities and QCD constraints at high densities.
In the intermediate region, the EoS is constrained by results for the symmetry energy and its slope at nuclear saturation, which are related to the pressure and energy per particle of neutron matter at that density. Additionally, the equation of state is also constrained by mechanical stability, causality and thermodynamic consistency (see Ref.~\cite{Alarcon:2022vtn} for more details).

\begin{figure*}[!t]
  \centering
      \includegraphics[width=0.49\textwidth]{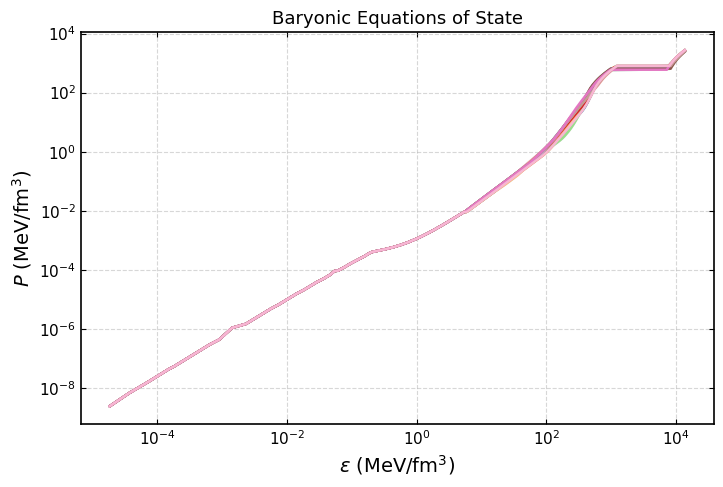} 
      \includegraphics[width=0.49\textwidth]{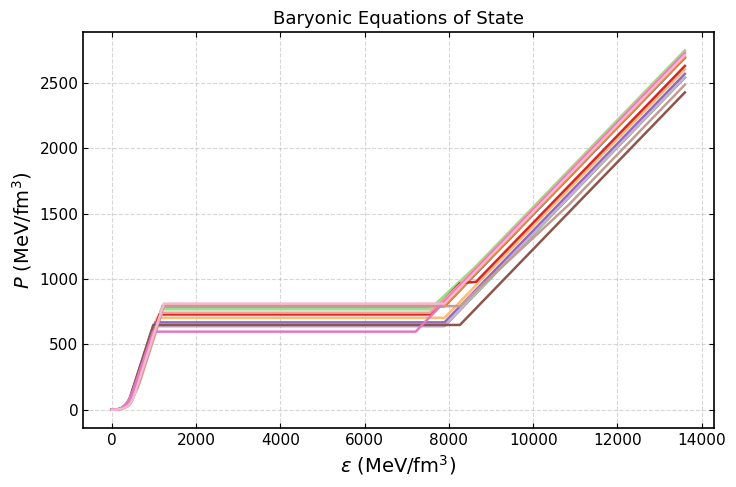} 
  \caption{Plot of the baryonic equations of state used in this work on logarithmic (left) and linear (right) scales.}
  \label{Fig:Baryonic_EoS}
\end{figure*}

\begin{figure}[h!]
  \centering
      \includegraphics[width=0.49\textwidth]{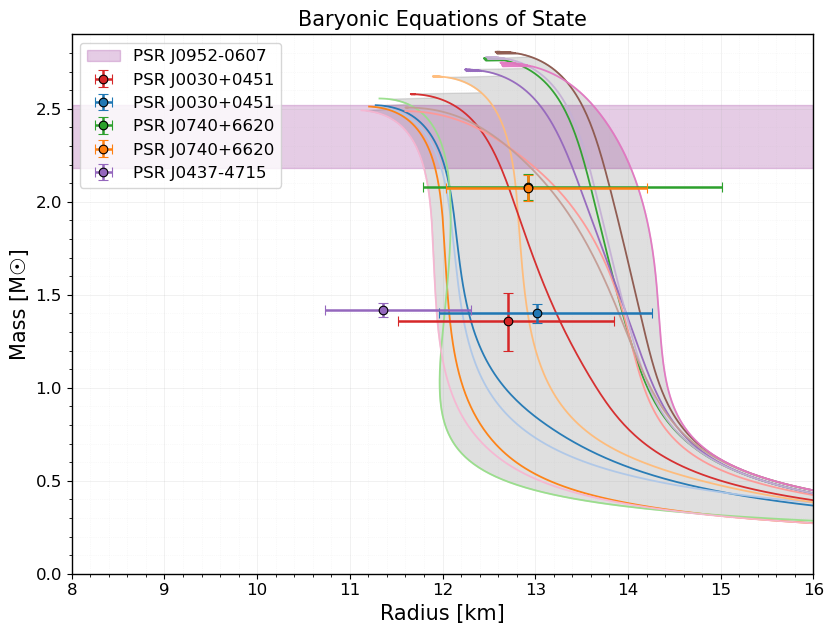} 
  \caption{Mass-radius diagrams generated from the EoSs used in this work. The neutron-star masses and radii are taken from \cite{Romani:2022jhd} (PSR J0952-0607), \cite{Riley:2019yda,Miller:2019cac} (PSR J0030+0451), \cite{Dittmann:2024mbo,Salmi:2024aum} (PSR J0740+6620), and \cite{Choudhury:2024xbk} (PSR J0437-4715).}
  \label{Fig:Baryonic}
\end{figure}

To be specific, we use a set of 14 EoSs calculated in the same way as was done in Ref.~\cite{Alarcon:2024hlj}, but without considering the HESS J1731-347 \cite{2022NatAs...6.1444D} measurement. The EoSs that we use in this work also include the effect of a crust, which has been incorporated in the way that was explained in Ref.~\cite{Alarcon:2025qmz}.
In Figs.~\ref{Fig:Baryonic_EoS} and \ref{Fig:Baryonic} we plot the set of equations of state used in this work and the mass-radius diagrams they generate, respectively.

\subsection{Dark Matter EoS}

In this work, we analyze the effects of DM on the mass-radius relation of an NS.
The model used for DM is a relativistic non-interacting spin-1/2 fermion with only one species in the compact object.
We consider a non-interacting fermion because it moves the mass-radius diagram toward the CO posteriors, whereas the interacting (repulsive) Fermi gas moves it away from them \cite{Miao:2022rqj}.

The EoS of a free Fermi gas can be parameterized in terms of the Fermi momentum ($k_F$) in the following form

\begin{figure}[h!]
  \centering
      \includegraphics[width=0.49\textwidth]{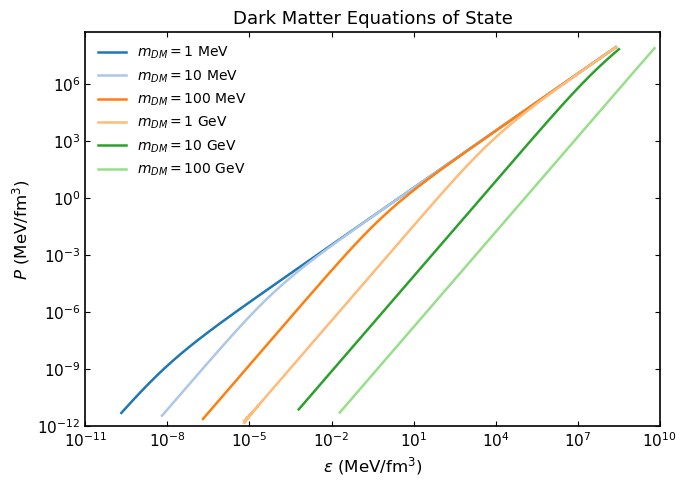} 
  \caption{DM equations of state used in this work.}
  \label{Fig:DM_EoS}
\end{figure}

{\small
\begin{gather}
    \begin{split}
        \varepsilon &= \frac{1}{\pi^2}\int_0^{k_F}k^2\sqrt{k^2+m_f^2}\ dk= \\
        &=\frac{1}{8\pi^2}\bigg[k_F\sqrt{k_F^2+m_f^2}(2k_F^2+m_f^2) -m_f^4\ln{\bigg(\frac{k_F+\sqrt{k_F^2+m_f^2}}{m_f}\bigg)} \bigg]
    \end{split} \\
    \begin{split}
        P &=\frac{1}{3\pi^2}\int_0^{k_F}\frac{k^4}{\sqrt{k^2+m_f^2}}\ dk \frac{1}{24\pi^2} \\
        &\times \bigg[k_F\sqrt{k_F^2+m_f^2}(2k_F^2-3m_f^2)
        +3m_f^4\ln{\bigg(\frac{k_F+\sqrt{k_F^2+m_f^2}}{m_f}\bigg)} \bigg]
    \end{split} 
    \label{Gas de Fermi libre}
\end{gather}
}
where we can express the Fermi momentum in terms of the number density ($n$) $k_F^3=3\pi^2n $ so that the only free parameter is the fermion mass $m_f$. In this study the fermion mass is simply the proposed DM mass, so $m_{DM}\equiv m_f$. For a representative range of possible DM candidates we choose six different DM particle masses $m_{DM}=\{1,10,10^2,10^3,10^4,10^5\}$~MeV. 
In Fig.~\ref{Fig:DM_EoS} we show the EoSs used for the DM component. 
Note that, depending on the DM mass, one needs to cover different ranges in $P$ and $\varepsilon$ to calculate Dark Matter Admixed Neutron Stars (DANS) with masses up to $\sim 2.7M_\odot$.

\subsection{Hydrostatic Equilibrium}
\label{Sec:Hydrostatic_equilibrium}

To study a star composed of an admixture of dark and baryonic matter, we assume that the two components interact only gravitationally.
For DM in the GeV mass range or above, this premise is experimentally justified because the cross section for nongravitational interactions between the two components is constrained to be very small, $\sigma_{\chi N}\sim10^{-45}\text{~cm}^2 \ll \sigma_{nucl}\sim10^{-24}\text{~cm}^2$~\cite{Grippa:2024ach}, where $\sigma_{\chi N}$ denotes the DM--nucleon cross section and $\sigma_{nucl}$ the nuclear cross section.
For light DM ($m_\chi\lesssim 100$~MeV), the estimated cross sections satisfy $\sigma_{\chi N}\lesssim 10^{-32}\text{~cm}^2$~\cite{COSINUS:2026acs,Cox:2024rew,TerolCalvo:2026xee}, which is also well below $\sigma_{nucl}$.

Under these assumptions, we can apply the two-fluid formulation of the Tolman--Oppenheimer--Volkoff (TOV) equations, in which each component is treated as a perfect fluid~\cite{Sandin:2008db} and its energy--momentum tensor is conserved separately. In this approach, the hydrostatic equilibrium of the system can be expressed as follows (in geometric units, $G=c=1$):
\begin{gather} \label{TOVpressure}
    \frac{dP_{BM}}{dr} = -(P_{BM}+\varepsilon_{BM})\frac{4\pi r^3(P_{BM} + P_{DM})+M(r)}{r(r-2M(r))} \\
    \frac{dP_{DM}}{dr} = -(P_{DM} +\varepsilon_{DM})\frac{4\pi r^3(P_{BM} + P_{DM})+M(r)}{r(r-2M(r))}
\end{gather}
where $r$ is the radial coordinate, $P_i$ is the pressure of each fluid, and $\varepsilon_i$ is its energy density. The total mass of the star and the mass of each component are given by
\begin{equation} \label{mass}
    M(r) = \sum_iM_i(r) \qquad \frac{dM_i(r)}{dr} = 4\pi\varepsilon_i(r)r^2
\end{equation}
where $i=BM,DM$.

 
Eqs.~\eqref{TOVpressure}--\eqref{mass} require both equations of state as input. To solve this system of coupled differential equations, we assume that the BM and DM masses inside a sphere of zero radius (the center of the star) are zero, $M_{DM}(0)=M_{BM}(0)=0$, and provide a set of initial central pressures. For the baryonic component, we consider the range $P_{BM}^c \in (757.5 \times 10^{-5},2.147\times 10^{3})$~MeV/fm$^3$. For the dark matter component, we use different ranges depending on the DM particle mass so that we can reach the desired DM fractions. The full range of DM central pressures is contained within $P_{DM}^c \in (7.575 \times 10^{-5},5.88\times10^{10})$~MeV/fm$^3$. We chose these central-pressure ranges because they cover the full range of stable configurations in the mass region of interest for our study.

For each pair of initial pressure conditions, we use the fourth-order Runge--Kutta method to solve the coupled equations outward from the core and obtain the final radius and mass of the star.
Once the pressure of any of the two fluids reaches a value below $P_{\text{min}} \sim 3\times 10^{-9} $~MeV/fm$^3$ (the minimum pressure of the crust EoS of the baryonic component) for a given distance, we consider this distance to be the radius of that component. 
Then, the calculation continues with only the other component until it reaches a value below $P_{\text{min}}$ as well. 
This allows for two possible scenarios for the star: a dark matter core or a dark matter halo expanding beyond the baryonic star. 

\begin{figure*}[t!]
    \centering
    \includegraphics[width=0.6\textwidth]{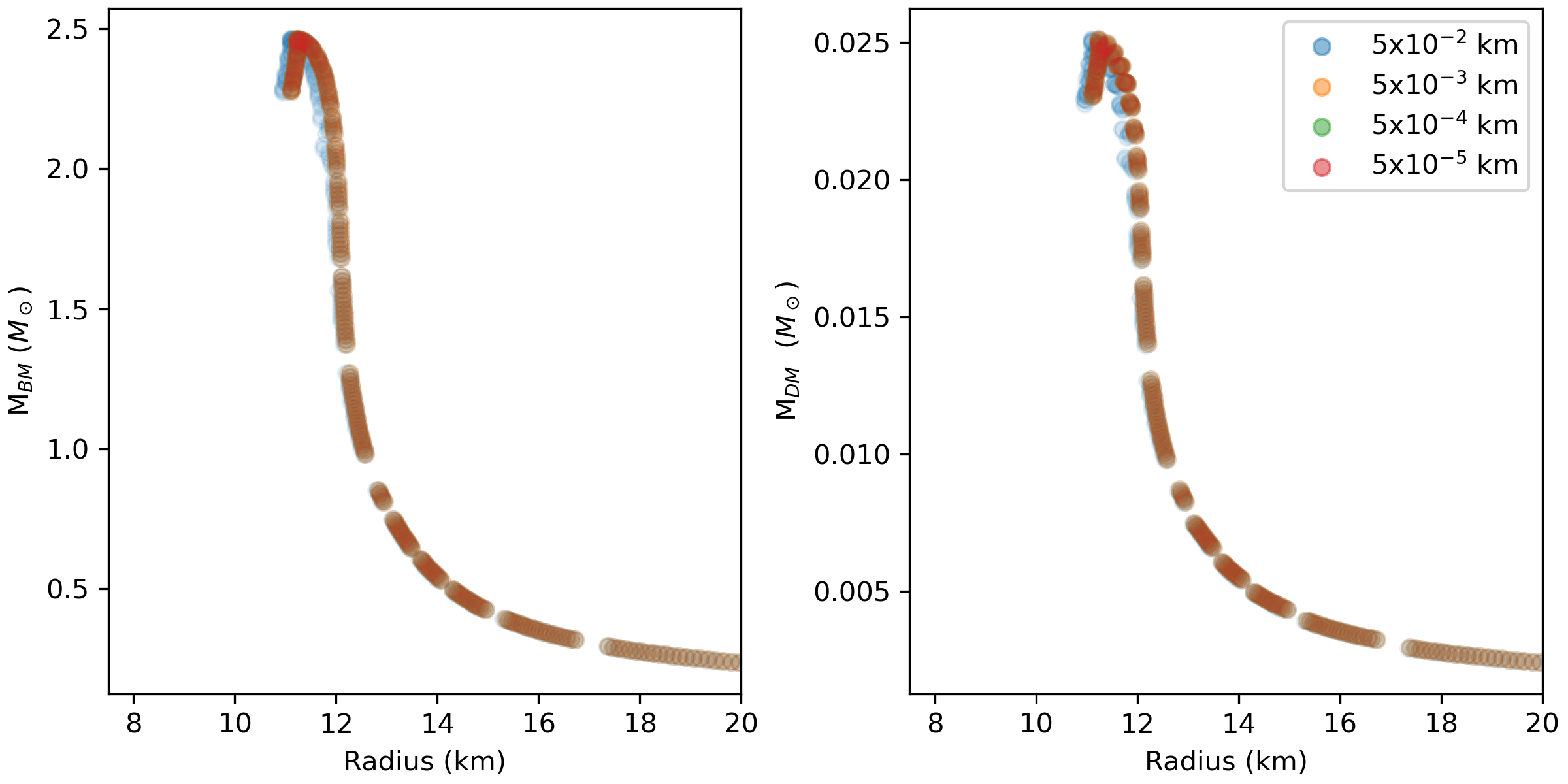}
    \caption{Baryonic and dark matter mass content as a function of the baryonic radius for the case of $m_{DM}=1$~GeV. Different colors represent different integration step sizes.}
    \label{fig:step comparison}
    
\end{figure*}

We performed a convergence study to check whether the reported results are independent of the integration step. This is illustrated in Fig.~\ref{fig:step comparison} for $m_{DM}=1$~GeV, which shows that integration steps of $5\times10^{-3}$~km and below yield the same mass-radius results independently of the integration step for both the baryonic and dark matter components. Similar results are obtained for the remaining DM masses. The results presented in this work are obtained with an integration step of $5\times10^{-5}$~km.

\subsection{Stability Analysis}
\label{Sec:Stability_Analysis}

To analyze the stability of the resulting star, we note that, for a dark matter admixed neutron star (DANS), the TOV equations do not yield a single mass-radius curve but instead multiple stellar configurations, one for each combination of central pressures.
One can therefore trace different curves based on different criteria, and curves for a fixed DM fraction differ from curves for a fixed ratio of central pressures.
Consequently, one cannot use the maximum-mass criterion for the onset of instability, as is typically done for one-fluid configurations. In this work, we follow the analysis presented in \cite{Hippert:2022snq}, which we briefly explain here.

In a two-fluid star, the change between stable and unstable regions occurs when the number of particles for each component remains stationary under variations in either of the energy densities  $\varepsilon_i^c$
\small
\begin{equation}
\label{stability}
    \begin{pmatrix}
        \delta N_{BM}\\
        \delta N_{DM}
    \end{pmatrix}= 
    \begin{pmatrix}
        \partial N_{BM}/\partial\varepsilon_{BM}^c & \partial N_{BM}/\partial\varepsilon_{DM}^c \\
        \partial N_{DM}/\partial\varepsilon_{BM}^c & \partial N_{DM}/\partial\varepsilon_{DM}^c
    \end{pmatrix}
    \begin{pmatrix}
        \delta\varepsilon_{BM}^c \\
        \delta\varepsilon_{DM}^c
    \end{pmatrix} = 0
\end{equation}
\normalsize
where $N$ denotes the number of particles and $\delta$ denotes small perturbations. To obtain nontrivial solutions to this equation, the determinant of the matrix must be zero. In the single-fluid scenario, this determinant condition yields $\partial N/\partial \varepsilon^c = 0$. Provided that the TOV equations are satisfied, this condition is equivalent to the established criterion $\partial M/\partial \varepsilon^c = 0$, which signifies the onset of instability. When computing $\partial N_i / \partial \varepsilon^c_j$, the central energy density of the other fluid is held fixed.
The matrix of Eq. \eqref{stability} can be diagonalized so that
\begin{equation}
    \begin{pmatrix}
        \delta N_A \\
        \delta N_B
    \end{pmatrix}=
    \begin{pmatrix}
        \kappa_A & 0 \\
        0 & \kappa_B
    \end{pmatrix}
    \begin{pmatrix}
        \delta\varepsilon_c^A \\
        \delta\varepsilon_c^B
    \end{pmatrix}
\end{equation}
where $N_A$ and $N_B$ are linear combinations of $N_{DM} $ and $N_{BM} $. Thus, stable configurations are only permissible when both eigenvalues are positive. For each fluid, it is feasible to compute the number of particles with the TOV equations.
\begin{equation}
    \frac{dN_i}{dr} =  \frac{4\pi n_ir^2}{\sqrt{1-2M(r)/r}}
\end{equation}

\subsection{DM Content Analysis}
In order to analyze the DANS hypothesis, we perform a Bayesian analysis of the mass-radius diagrams obtained with the different DM fractions considered in this work.
To obtain the observational likelihoods needed in the analysis, we use the published posteriors for HESS J1731-347 \cite{Doroshenko2022Zenodo} and PSR J1231-1411 \cite{Qi_2025}, while for XTE J1814-338 we obtained these functions through private communication with the authors.
\\
These posteriors are related to the likelihood we need to evaluate the Bayesian evidence $Z$,
\\
\begin{align}
   p(M,R|d) \propto \mathcal{L}(d|M,R)\pi(M,R)
\end{align}
where $\pi(M,R)$ and $p(M,R|d)$ are the prior used in the observational analysis and the posterior provided, respectively, and $\mathcal{L}(d|M,R)$ is the likelihood required for the Bayesian analysis.
The likelihood used in our work is a marginalized likelihood over the nuisance parameters of the observational analyses, and will be determined from the posterior and the knowledge of the observational prior in $M$ and $R$ used by each analysis.
\\
For the construction of the likelihood used to evaluate the HESS J1731-347 Bayesian evidence under the DANS hypothesis, we consider the posterior samples that exclude the additional EoS and astrophysical constraints.
We reconstructed a two-dimensional density from the full Markov Chain Monte Carlo (MCMC) and used the same Gaussian-kernel procedure employed by the authors of the publication.
Since the analysis states that flat priors were used for the mass and radius, we consider a uniform prior in these variables to reconstruct the likelihood. 
So, for HESS J1731-347, this translates to the following likelihood,
\\
\begin{align}
    \mathcal{L}_{\text{HESS}}(d|M,R) \propto p_{HESS}(M,R|d)
\end{align}
\\
In the case of PSR J1231-1411, we used the posterior obtained with the STS-PST result.
This analysis employed a normal distribution for the gravitational-mass prior, with parameters $\mu_M = 1 M_\sun$ and $\sigma_M = 0.93 M_\sun$, over the support $1 M_\sun \leq M \leq 2 M_\sun$. For the radius, the authors considered a uniform prior with support $8~\text{km} \leq R \leq 16~\text{km}$.
We compute the normalized truncated Gaussian mass prior $\pi_M(M)$ and the uniform distribution for the radius prior $\pi_R(R)$ to reconstruct the likelihood as
\\
\begin{align}
    \mathcal{L}_{\text{PSR}}(d|M,R) \propto \frac{p_{PSR}(M,R|d)}{\pi_M(M) \pi_R(R)}
\end{align}
\\
For the third compact object, XTE J1814-338, we use the posterior provided by the authors of the publication.
Since the analysis considers independent uniform priors for the gravitational mass and radius in the ranges $1 M_\sun \leq M \leq 3 M_\sun$ and $3 G M_\sun/c^2 \leq R/\text{km} \leq 16$, the required likelihood is proportional to the provided posterior:
\\
\begin{align}
    \mathcal{L}_{XTE}(d|M,R) \propto p_{XTE}(M,R|d)
\end{align}
\\
\\
Once we have the required likelihoods for the three compact objects, we evaluate the Bayesian evidence of the different mass-radius diagrams generated under the DANS hypothesis for different DM fractions and masses.
For each baryonic EoS with a given DM fraction and particle mass, we calculate the Bayesian evidence of the stable sequence by integrating the likelihood of a given observation over the curve as follows:
\begin{align}\label{Eq:Z_definition}
\begin{split}
Z^{DANS}(\alpha,m_{DM},f_{DM})=&\int_{M_{min}}^{M_{max}}\mathcal{L}\left[M,R(M,\alpha,m_{DM},f_{DM})\right] \\
&\times \pi(M| \alpha,m_{DM},f_{DM})\,dM  
\end{split}  
\end{align}
where $\alpha$ denotes the baryonic EoS used, and $M_{min}$ and $M_{max}$ correspond to the minimum and maximum masses of the stable sequence, respectively.
If the stable sequence has a mass gap, the previous integral is split into a sum of integrals over the stable sequences, so unstable solutions do not contribute to Eq.~\eqref{Eq:Z_definition}.
A uniform prior is used in the mass integration because there is no evidence favoring particular mass configurations in a DANS scenario. This prior is defined as $\pi(M| \alpha,m_{DM},f_{DM}) = (M_{max}-M_{min})^{-1}$. We also marginalize over the baryonic equations of state as follows:
\begin{align}
Z^{DANS}(m_{DM},f_{DM}) = \frac{1}{14}\sum_{\alpha=1}^{14} Z^{DANS}(\alpha,m_{DM},f_{DM})
\end{align}  
in order to obtain the evidence of the DANS scenario with a given $m_{DM}$ and $f_{DM}$ for a certain observation. 

The same procedure is applied to the pure baryonic solutions, marginalizing over the 14 baryonic EoS, in order to obtain the evidence of the purely baryonic NS scenario (PBNS),

\begin{align}
\begin{split}
&Z^{PBNS}(\alpha)=\int_{M_{min}}^{M_{max}}\mathcal{L}\left[M,R(M,\alpha)\right]\pi(M|\alpha)\,dM \\   
& Z^{PBNS} = \frac{1}{14}\sum_{\alpha=1}^{14} Z^{PBNS}(\alpha)
\end{split}
\end{align}

Using the evidence computed for the DANS and PBNS hypotheses, we evaluate the Bayes factor for given values of $m_{DM}$ and $f_{DM}$ by taking the ratio

\begin{align}
B_{CO}(m_{DM},f_{DM}) = \frac{Z_{CO}^{DANS}(m_{DM},f_{DM})}{Z_{CO}^{PBNS}},
\end{align}

where $CO$ refers to one of the three compact objects, and can take the labels HESS, PSR or XTE.
Its value allows to analyze which ($m_{DM}$, $f_{DM}$) combination is favored or disfavored over the PBNS hypothesis for each compact object. 

It is also useful to marginalize over the DM fraction to estimate which DM mass most strongly favors the DANS hypothesis for each compact object.
This is done for the three COs separately and together (joint analysis).
With this aim, we perform the marginalization with a uniform prior in $f_{DM}$ in the following way
\begin{align}
Z^{DANS}(m_{DM})= \frac{1}{10}\sum_{f_{DM}} Z^{DANS}(m_{DM},f_{DM}),
\end{align}  
\\
since we have ten possible values for $f_{DM}$ for each DM mass. In this case, the Bayes factor is calculated as follows 
\begin{align}
B_{CO}(m_{DM}) = \frac{Z_{CO}^{DANS}(m_{DM})}{Z_{CO}^{PBNS}}
\end{align}  

for the analysis of a single CO.
We can combine the information of the three compact objects to infer the most likely DM mass among the ones studied in this work under the DANS hypothesis. 
In this joint analysis, we first calculate $Z^{DANS}(\alpha,m_{DM},f_{DM})$ [Eq.~\eqref{Eq:Z_definition}] and marginalize over the DM fractions for a given EoS and compact object,

\begin{align}
 Z_{CO}^{DANS}(\alpha,m_{DM})= \frac{1}{10}\sum_{f_{DM}}Z_{CO}^{DANS}(\alpha,m_{DM},f_{DM})
\end{align} 

and construct the joint evidence as the product of $Z_{CO}^{DANS}(\alpha,m_{DM})$ for the three compact objects:

\begin{align}
\begin{split}
 &Z_{joint}^{DANS}(\alpha,m_{DM})= Z_{HESS}^{DANS}(\alpha,m_{DM})\\
 &\times Z_{PSR}^{DANS}(\alpha,m_{DM})\ Z_{XTE}^{DANS}(\alpha,m_{DM}).
\end{split}
\end{align} 

This evidence is then marginalized over the EoS in order to get the evidence under the DANS hypothesis for a given DM mass 

\begin{align}
 Z_{joint}^{DANS}(m_{DM}) = \frac{1}{14}\sum_{\alpha=1}^{14} Z_{joint}^{DANS}(\alpha,m_{DM}).
\end{align}

The posterior probability of each DM mass is then calculated from $Z_{joint}^{DANS}(m_{DM})$ after normalization. The evidence for the PBNS hypothesis is calculated in the same way:

\begin{align}
\begin{split}
 &Z_{joint}^{PBNS}(\alpha) = Z_{HESS}^{PBNS}(\alpha)\ Z_{PSR}^{PBNS}(\alpha)\ Z_{XTE}^{PBNS}(\alpha)  \\
 &Z_{joint}^{PBNS} = \frac{1}{14}\sum_{\alpha=1}^{14} Z_{joint}^{PBNS}(\alpha),
\end{split}
\end{align}

and the Bayes factor of the joint analysis is obtained by taking the ratio

\begin{align}
B_{joint}(m_{DM}) = \frac{Z_{joint}^{DANS}(m_{DM})}{Z_{joint}^{PBNS}}.
\end{align}

The value of the Bayes factor indicates whether the DANS hypothesis is favored ($>1$) or disfavored ($<1$) over the pure baryonic hypothesis.

\section{\textbf{Results}}\label{Sec:Results}

Depending on the DM mass, the DM component is either completely contained within the baryonic radius (core configuration) or extends beyond it (halo configuration). It is therefore important to define the mass and radius shown in the M-R diagrams, especially for halo configurations.
The mass and radius measurements of PSR J1231-1411 and HESS J1731-347 were obtained through spectral modeling with XSPEC, whereas those of XTE J1814-338 were obtained through pulse-profile modeling using X-PSI.
In these works, the reported masses and radii of the COs are those that accurately reproduce the measured X-ray spectra.
For these reasons, all mass-radius figures plotted in this work show the baryonic radius $R_{BM}$ and the mass contained within it, $M_{BM}(R_{BM})+M_{DM}(R_{BM})$.
The DM fractions are calculated consistently with this choice, so that the fraction refers to the DM contained within the baryonic radius ($f_{DM}=M_{DM}(R_{BM})/[M_{BM}(R_{BM})+M_{DM}(R_{BM})]$). Note that, with this definition, COs with halo configurations may contain substantially more DM (including the halo) than the value of $f_{DM}$ suggests.

\begin{figure*}[!t]
    \centering
    \includegraphics[width=0.75\textwidth]{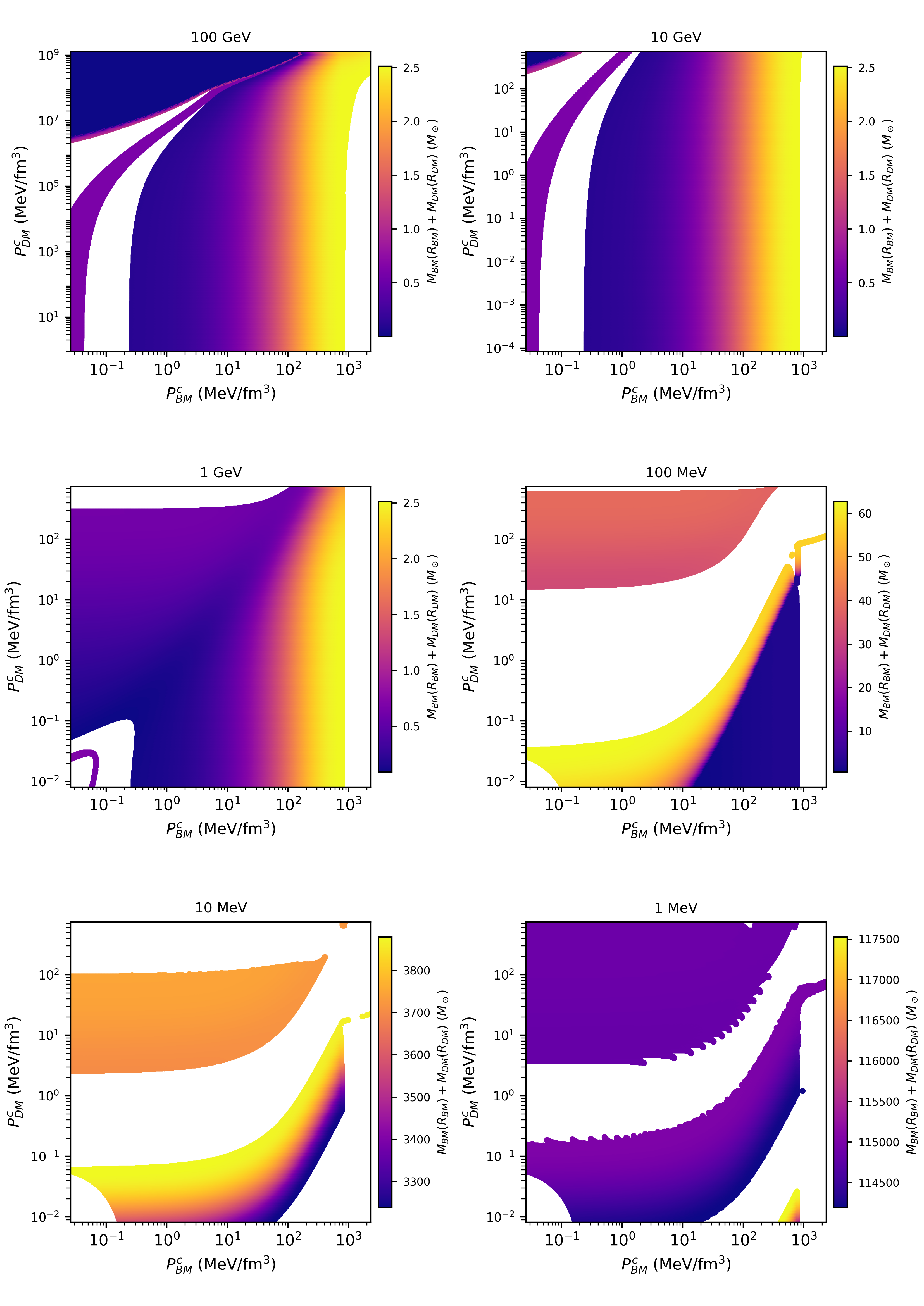}
    \caption{Results for the stability analysis with the different DM masses for one given representative baryonic EoS. The stable admixed configurations are shown colored. The unstable configurations appear in white. The total gravitational mass is shown as a function of both central pressures.}
    \label{Fig:stability_all_DM_mases_nuevo}
    
\end{figure*}

\begin{figure*}[t]
  \centering
      \includegraphics[width=0.8\textwidth]{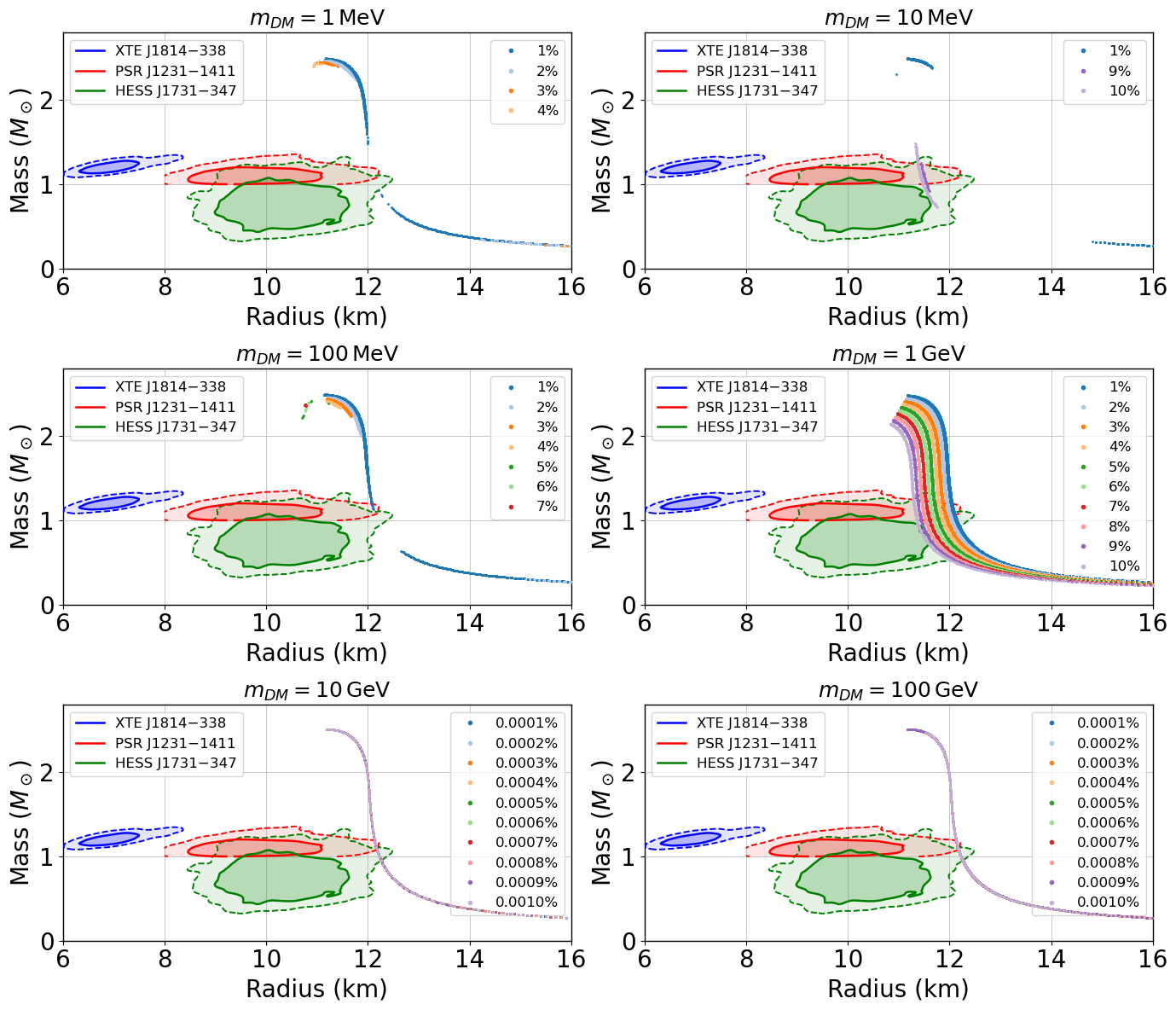} 
  \caption{In this figure, we show the effect of the different fractions and different DM particle mass on the mass-radius diagram for one of the baryonic EoS.
  The green, red and blue contours correspond to the HESS J1731-347, PSR J1231-1411 and XTE J1814-338 posteriors at 68\% (solid) and 95\% (dashed) credible region.}
  \label{1EoS varias masas}
\end{figure*}

We performed the stability analysis for the six DM particle masses by studying the baryonic and dark matter central pressures $P_{BM}^c$ and $P_{DM}^c$.
In order to analyze the DANS hypothesis, we consider only the mass-radius sequence that satisfies the stability condition described in Sec.~\ref{Sec:Stability_Analysis} for a given DM fraction. Figure~\ref{Fig:stability_all_DM_mases_nuevo} shows the combination of central pressures that give rise to stable configurations, and the resulting total gravitational mass of the compact object. 
For all DM masses, most configurations with a baryonic central pressure above the phase transition become unstable. Small DM masses all show similar results, which was expected due to the similar DM EoS for light DM. They show more unstable configurations than heavy DM masses. For the two lightest DM masses, the configurations with larger gravitational mass are those with high baryonic central pressure and low DM central pressure. We also observe that small DM masses give rise to halo configurations with unusually large total gravitational masses. This is discussed in more detail later in this section. For heavier DM masses, the stability condition is mainly dependent on the baryonic central pressure.
These results do not vary significantly between baryonic EoSs.

\begin{figure*}[t]
    \centering
    \includegraphics[width=0.45\linewidth]{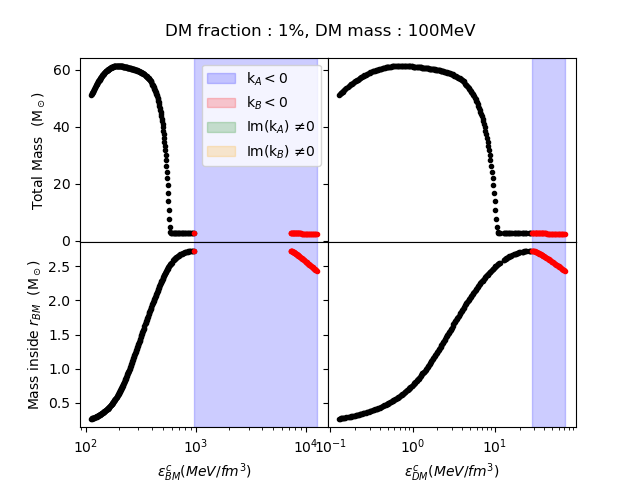}
    \includegraphics[width=0.45\linewidth]{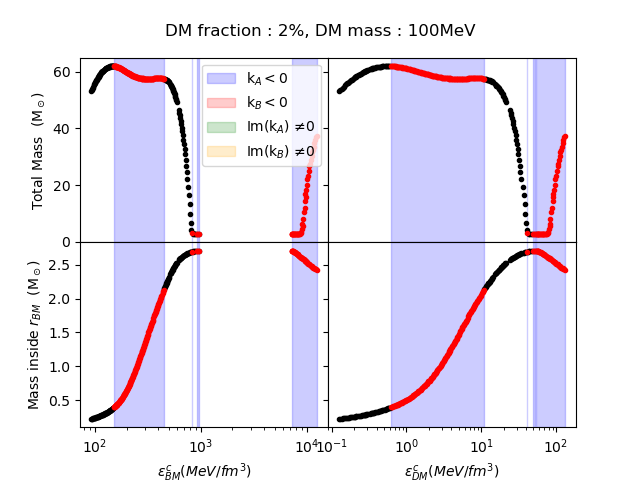}
    \caption{Stability analysis for a DM fraction of $1\%$ (left) and $2\%$ (right) for a DM mass of $100$~MeV and for an EoS that produces unstable solutions in the intermediate mass region.
    The black points show the stable solutions, while the red points show the unstable ones. Shaded areas of different colors show the eigenvalue condition that makes the configuration unstable.}
    \label{Fig:stability_additional_figure_100MeV}
\end{figure*}

\begin{figure*}[t]
    \centering
    \includegraphics[width=0.45\linewidth]{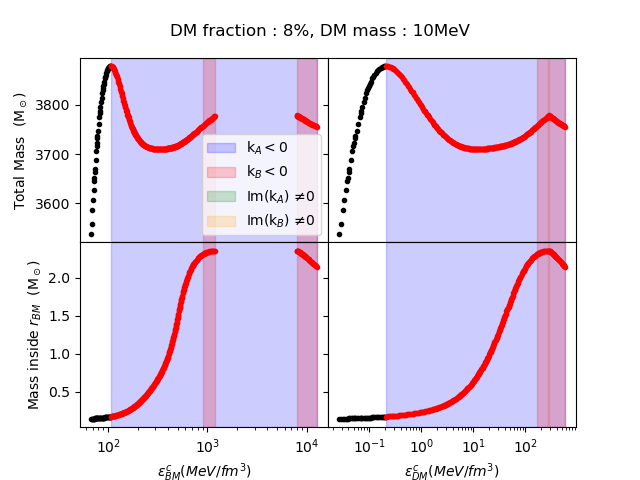}
    \includegraphics[width=0.45\linewidth]{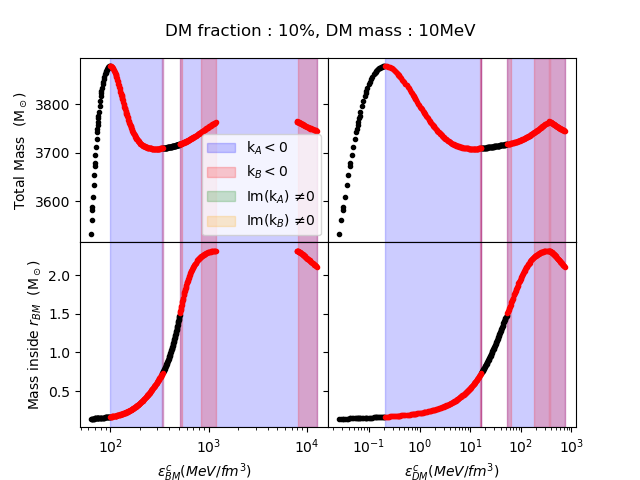}
    \caption{Stability analysis for a DM fraction of $8\%$ (left) and $10\%$ (right), for a DM mass of $10$~MeV and for an EoS that produces some stable solutions in the intermediate mass region.
    The black points show the stable solutions, while the red points show the unstable ones. Shaded areas of different colors show the eigenvalue condition that makes the configuration unstable.}
    \label{Fig:stability_additional_figure_10MeV}
\end{figure*}

\begin{figure*}[!t]
\centering

\begin{subfigure}[t]{0.45\textwidth}
\centering
\includegraphics[width=\textwidth]{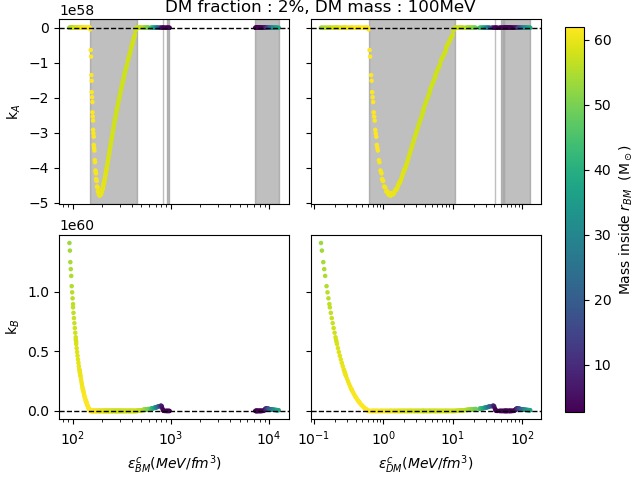}
\end{subfigure}%
\hfill
\begin{subfigure}[t]{0.45\textwidth}
\centering
\includegraphics[width=\textwidth]{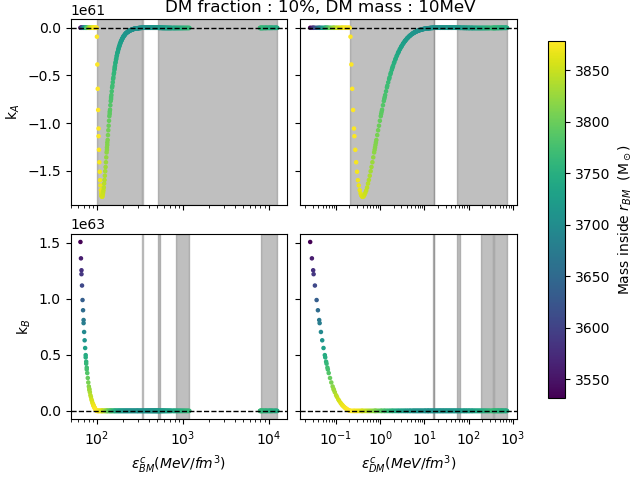}

\end{subfigure}

\begin{subfigure}[t]{0.45\textwidth}
\centering
\includegraphics[width=\textwidth]{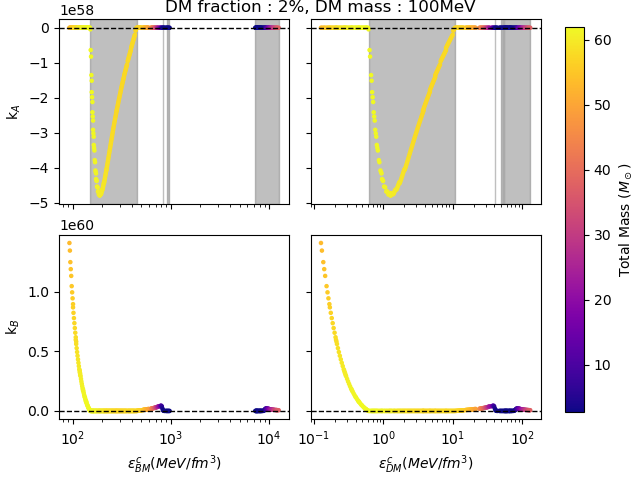}

\end{subfigure}%
\hfill
\begin{subfigure}[t]{0.45\textwidth}
\centering
\includegraphics[width=\textwidth]{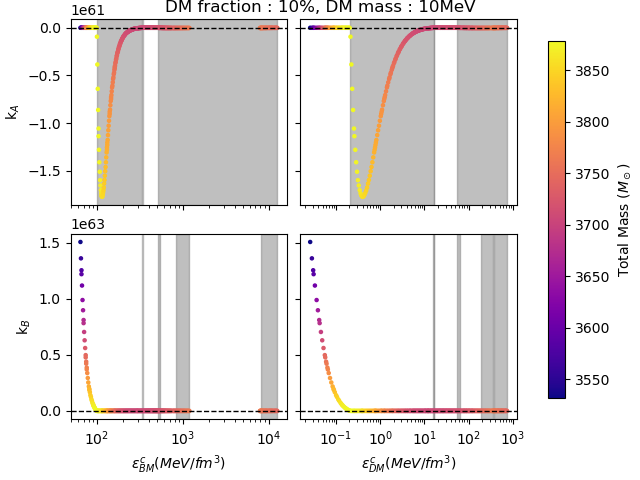}

\end{subfigure}

\caption{Dependence of the eigenvalues on the central energy densities of both components for a given baryonic EoS. The left (right) figure corresponds to a DM mass of $100$~MeV ($10$~MeV) and a DM fraction of $2\%$ ($10\%$). The colors in the top panels represent the mass contained within the baryonic radius, while those in the bottom panels represent the total gravitational mass. The shaded gray region appears where the shown eigenvalue is negative. The dashed line marks zero on the y-axis.}
\label{fig:kA kB}
\end{figure*}

We solve the TOV equations in the two-fluid approach described in Sec.~\ref{Sec:Hydrostatic_equilibrium} to obtain the stable M-R sequence of a DANS with a given DM mass and different DM fractions.
Fig.~\ref{1EoS varias masas} shows the results that we obtain for all DM masses and for a given baryonic EoS. In the Appendix~\ref{App:M-R_diagrams_all_EoS} we show the results for the M-R diagrams of stable configurations with all the baryonic EoSs for the different DM masses considered in this work.
The results obtained from these calculations allow us to distinguish three different regimes.

For the heaviest DM masses ($m_{DM}=100$ and $10$ GeV), we obtain core configurations, and DM fractions of the order of $10^{-4}-10^{-3}\%$ lead to stable results, while higher ones lead to unstable sequences. 
For this small DM content, the modification of the M-R relation is negligible with respect to the pure baryonic scenario, as seen in Fig.~\ref{1EoS varias masas}. This is shown even more clearly in the Fig.~\ref{Fig:100GeV} of the Appendix.
Since the DM component does not have a noticeable impact on the M-R curves of the stable solutions for these masses, the level of agreement with the three compact objects is determined almost entirely by the baryonic EoS, and it is not possible to infer the presence of a DM component from the M-R relation.

In the case of a DM mass of $1$~GeV, we observe a completely different behavior.
For this mass we obtain stable core configurations for DM fractions between $1\%$ and $10\%$. 
These fractions are high enough to modify significantly the M-R relation with respect to the pure baryonic scenario.
As the DM fraction increases, the M-R curves move toward lower masses and smaller radii. 
The magnitude of this shift is approximately constant for the $1\%$ step used for the DM fraction.
This behavior shows that an admixed DM component can explain the low-mass compact objects that are difficult to reproduce with a purely baryonic EoS. 
In particular, for HESS J1731-347 and PSR J1231-1411 the highest DM fractions move the M-R curves to the 68\% credible region, as shown in Fig.~\ref{1EoS varias masas}. 
However, the M-R curves obtained for this mass are still far away from the 95\% credible region of the XTE J1814-338 compact object. 

An entirely different scenario for the COs emerges when $m_{DM}\leq 100$~MeV.
In this case, the DM component extends beyond the baryonic radius, forming a halo surrounding the baryonic object.
This halo may contain a very large amount of mass, reaching several tens of solar masses for $100$~MeV and up to several thousand solar masses for $10$ and $1$~MeV (see Fig. \ref{Fig:stability_all_DM_mases_nuevo}). 
As explained at the beginning of this section, we use the DM mass inside the baryonic radius to calculate the mass of the compact object, while the radius plotted in the M-R curves corresponds to the baryonic one. 
Therefore, in the analysis performed here, the halo affects only the stability of the CO, not the mass or radius shown in the figures.
For $m_{DM}=100$~MeV, the emerging halo introduces unstable regions within the mass range of the three COs for DM fractions $\geq 2\%$, while for $f_{DM}=1\%$ there are still stable solutions in that mass region for most of the baryonic EoSs employed. Consequently, the resulting M-R curves show a gap for $f_{DM}\geq 2\%$, as seen in Fig.~\ref{Fig:1GeV} (right panel).
This gap is also obtained for the lightest DM mass ($m_{DM}=1$~MeV), while for $m_{DM}=10$~MeV there are a few stable solutions for fractions of approximately $9\%$ and $10\%$.
In order to understand the cause of these unstable solutions, in Figs. \ref{Fig:stability_additional_figure_100MeV} and \ref{Fig:stability_additional_figure_10MeV} we show both, the gravitational mass, and the mass inside the baryonic radius as a function of both energy densities.

\begin{table*}[ht!]
\centering
\begin{tabular*}{\textwidth}{@{\extracolsep{\fill}}c|cccc||c|cc}
\hline\hline
$f_{DM}$
& 1 MeV & 10 MeV & 100 MeV & 1 GeV
&
$f_{DM}$
& 10 GeV & 100 GeV
\\
\hline
1\% & $0$ & $0$ & $0.0256$ & $1.25$ & 0.0001\% & $1$ & $1$ \\
2\% & $0$ & $0$ & $0$ & $1.6$ & 0.0002\% & $1$ & $1$ \\
3\% & $0$ & $0$ & $0$ & $2.05$ & 0.0003\% & $1$ & $1$ \\
4\% & $0$ & $0$ & $0$ & $2.47$ & 0.0004\% & $1$ & $1$ \\
5\% & $0$ & $0$ & $0$ & $2.95$ & 0.0005\% & $1$ & $1$ \\
6\% & $0$ & $0$ & $0$ & $3.48$ & 0.0006\% & $1$ & $1$ \\
7\% & $0$ & $0$ & $0$ & $3.99$ & 0.0007\% & $1$ & $1$ \\
8\% & $0$ & $0$ & $0$ & $4.47$ & 0.0008\% & $1$ & $1$ \\
9\% & $0$ & $0.708$ & $0$ & $5.11$ & 0.0009\% & $1$ & $1$ \\
10\% & $0$ & $3.47$ & $0$ & $5.77$ & 0.0010\% & $1$ & $1$ \\
\hline\hline
\end{tabular*}
\caption{Bayes factors $B_{CO}(m_{DM},f_{DM})$
for HESS J1731-347, before marginalization over $f_{DM}$.
The first five columns corresponds to $f_{DM}=1\%-10\%$ and
$m_{DM}=1~\mathrm{MeV}-1~\mathrm{GeV}$, while the last three columns
corresponds to $f_{DM}=0.0001\%-0.0010\%$ and
$m_{DM}=10-100~\mathrm{GeV}$. Entries with $0$ correspond to negligible values.
\label{Tab:Bayes_HESS_mDM_fDM}}
\end{table*}

\begin{table*}[ht!]
\centering
\begin{tabular*}{\textwidth}{@{\extracolsep{\fill}}c|cccc||c|cc}
\hline\hline
$f_{DM}$
& 1 MeV & 10 MeV & 100 MeV & 1 GeV
&
$f_{DM}$
& 10 GeV & 100 GeV
\\
\hline
1\% & $3.51\times 10^{-11}$ & $0$ & $0.0438$ & $1.20$ & 0.0001\% & $1$ & $1$ \\
2\% & $0$ & $0$ & $0$ & $1.43$ & 0.0002\% & $1$ & $1$ \\
3\% & $0$ & $0$ & $0$ & $1.68$ & 0.0003\% & $1$ & $1$ \\
4\% & $0$ & $0$ & $0$ & $1.95$ & 0.0004\% & $1$ & $1$ \\
5\% & $0$ & $0$ & $0$ & $2.25$ & 0.0005\% & $1$ & $1$ \\
6\% & $0$ & $0$ & $0$ & $2.58$ & 0.0006\% & $1$ & $1$ \\
7\% & $0$ & $0$ & $0$ & $2.94$ & 0.0007\% & $1$ & $1$ \\
8\% & $0$ & $0$ & $0$ & $3.34$ & 0.0008\% & $1$ & $1$ \\
9\% & $0$ & $1.22$ & $0$ & $3.78$ & 0.0009\% & $1$ & $1$ \\
10\% & $0$ & $3.26$ & $0$ & $4.27$ & 0.0010\% & $1$ & $1$ \\
\hline\hline
\end{tabular*}
\caption{Bayes factors $B_{CO}(m_{DM},f_{DM})$
for PSR J1231-1411, before marginalization over $f_{DM}$.
The first five columns corresponds to $f_{DM}=1\%-10\%$ and
$m_{DM}=1~\mathrm{MeV}-1~\mathrm{GeV}$, while the three right columns
corresponds to $f_{DM}=0.0001\%-0.0010\%$ and
$m_{DM}=10-100~\mathrm{GeV}$. Entries with $0$ correspond to negligible values.
\label{Tab:Bayes_PSR_mDM_fDM}}
\end{table*}

\begin{table*}[ht!]
\centering
\begin{tabular*}{\textwidth}{@{\extracolsep{\fill}}c|cccc||c|cc}
\hline\hline
$f_{DM}$
& 1 MeV & 10 MeV & 100 MeV & 1 GeV
&
$f_{DM}$
& 10 GeV & 100 GeV
\\
\hline
1\% & $0.116$ & $0$ & $1.58$ & $1.79$ & 0.0001\% & $1$ & $1$ \\
2\% & $0$ & $0$ & $7.04\times 10^{-3}$ & $6.48$ & 0.0002\% & $1$ & $1$ \\
3\% & $0$ & $0$ & $0$ & $16.3$ & 0.0003\% & $1$ & $1$ \\
4\% & $0$ & $0$ & $0$ & $29.4$ & 0.0004\% & $1$ & $1$ \\
5\% & $0$ & $0$ & $0$ & $45.9$ & 0.0005\% & $1$ & $1$ \\
6\% & $0$ & $0$ & $0$ & $49.5$ & 0.0006\% & $1$ & $1$ \\
7\% & $0$ & $0$ & $0$ & $31.0$ & 0.0007\% & $1$ & $1$ \\
8\% & $0$ & $0$ & $0$ & $14.6$ & 0.0008\% & $1$ & $1$ \\
9\% & $0$ & $0$ & $0$ & $9.16$ & 0.0009\% & $1$ & $1$ \\
10\% & $0$ & $0$ & $0$ & $8.53$ & 0.0010\% & $1$ & $1$ \\
\hline\hline
\end{tabular*}
\caption{Bayes factors $B_{CO}(m_{DM},f_{DM})$
for XTE J1814-338, before marginalization over $f_{DM}$.
The five first columns corresponds to $f_{DM}=1\%-10\%$ and
$m_{DM}=1~\mathrm{MeV}-1~\mathrm{GeV}$, while the three right columns
corresponds to $f_{DM}=0.0001\%-0.0010\%$ and
$m_{DM}=10-100~\mathrm{GeV}$. Entries with $0$ correspond to negligible values.
\label{Tab:Bayes_XTE_mDM_fDM}}
\end{table*}

\begin{table*}[ht!]
    \centering
    Bayes factors ($B_{CO}$)\\ 
    \begin{tabular*}{\textwidth}{@{\extracolsep{\fill}}c|cccccc}
     \hline\hline
       & $1$~MeV  & $10$~MeV  & $100$~MeV & $1$~GeV & $10$~GeV & $100$~GeV \\
       \hline
      HESS J1731-347 & $0$ & $0.42$ & $0.003$ & $3.31$ & $1.00$ & $1.00$ \\
      PSR J1231-1411 & $0$ & $0.45$ & $4.37\times 10^{-3}$ & $2.54$ & $1.00$ & $1.00$ \\
      XTE J1814-338 & $0.01$ & $0$ & $0.16$ & 21.27 & 1.00 & 1.00 \\
      \hline

    \end{tabular*}
    \caption{Bayes factors for the different $m_{DM}$ studied in this work.\label{Tab:Bayes_factors_mDM_separated_analysis}}
\end{table*}


In Fig. \ref{Fig:stability_additional_figure_100MeV} we see that, for a DM fraction of $2\%$, the unstable configuration that we see in the M-R diagram corresponds to solutions with large total gravitational mass (including the halo). When one considers only the mass inside the baryonic radius, these unstable configurations lie in the range of $\sim 0.5M_\odot-2.1M_\odot$, giving rise to the gap we see in Figs.~\ref{1EoS varias masas} and \ref{Fig:1GeV}. This gap gets wider as the DM fraction grows.
These unstable configurations can be traced back to the DM component of the star, since they appear because of a change of sign in $\partial N_{DM}/\partial\varepsilon^c_{DM}$.
In the case of a DM mass of $10$~MeV, the opposite scenario occurs. We see how, for $f_{DM} = 8\%$, there is no stable solution for masses inside the baryonic radius $\gtrsim 0.1M_{\odot}$. However, when the DM fraction increases to $10\%$, stable solutions emerge in the range of $1M_{\odot} -1.5M_{\odot}$. This is explained by a change of sign of the eigenvalue $\kappa_A$, due to the derivate $\partial N_{DM}/\partial\varepsilon^c_{DM}$ becoming positive.
In Fig.~\ref{fig:kA kB} we show which mass configurations become unstable due to the change of sign of the eigenvalues.

\begin{table}[ht!]
    \centering
    Joint Bayes factor ($\log_{10}B_{joint}$)\\ 
    \resizebox{\columnwidth}{!}{%
    \begin{tabular}{cccccc}
     \hline\hline
        $1$~MeV  & $10$~MeV  & $100$~MeV & $1$~GeV & $10$~GeV & $100$~GeV \\
       \hline
       -- & $-12.30$ & $-13.20$ & $2.19$ & $3.60\times 10^{-5}$ & $3.60\times 10^{-5}$ \\
      \hline
    \end{tabular}%
    }
    \caption{Logarithm of the Bayes factor obtained for each $m_{DM}$ in the joint analysis of the three COs. For $m_{DM} = 1$~MeV the Bayes factor is $\sim 0$. \label{Tab:Bayes_factors_mDM_joint_analysis}}
\end{table}


So, in summary, our study reveals a transition between three different regimes, depending on the DM mass considered. 
For the heaviest DM masses ($100$ and $10$~GeV), the stable configurations correspond to a DM fraction so small that the DM component has no observable impact on the mass and radius of the compact object. 
When the DM mass is of the order of $1$~GeV, the DM begins to modify the M-R curves, keeping the stability properties of a pure baryonic NS. 
For lower DM masses ($\leq 100$~MeV), we see a transition to a DM halo configuration of the CO, where unstable branches arise in the mass regime of interest (considering the mass inside the baryonic radius).

We performed the Bayesian analysis explained in the previous section by computing the evidence of each stable sequence for given values of $m_{DM}$ and $f_{DM}$ and a given baryonic EoS.
After marginalizing over the EoS, we compute the Bayes factor to compare the DANS and pure-baryonic hypotheses for each combination of DM mass and DM fraction inside the CO.
The results are shown in Tables~\ref{Tab:Bayes_HESS_mDM_fDM}, \ref{Tab:Bayes_PSR_mDM_fDM} and \ref{Tab:Bayes_XTE_mDM_fDM}. 

They show that, for the CO HESS J1731-347, the Bayes factor grows monotonically with the DM fraction when $m_{DM} = 1$~GeV, maximizing the DANS hypothesis over the PBNS one for the maximum fraction considered here. 
However, for the light DM masses ($m_{DM} \leq 100$~MeV), the evidence of the DANS hypothesis becomes lower than the evidence of the PBNS hypothesis, giving Bayes factors lower than $1$.
The reason is the small number of stable solutions in the regions with non-negligible probability density for the three COs.
On the other hand, the curves generated with a DM mass of $m_{DM} = 10$~GeV and $m_{DM} = 100$~GeV  barely move with respect to the PBNS for these fractions, so the Bayes factor is $\approx 1$.
This situation is the same for the PSR J1231-1411 CO,
where the most likely fractions correspond to the highest values for $m_{DM} = 1$~GeV.

In the case of the XTE J1814-338 CO, we get a different scenario. 
For this CO, the M-R region of highest probability density lies far from all stable sequences under the DANS and PBNS hypotheses, and the regions that contribute most to the evaluated evidence correspond to the highest masses ($\gtrsim 2M_{\odot}$).
For a DM mass of $1$~GeV, this translates into an increasing Bayes factor up to $f_{DM}=6\%$, for which the stable sequence reaches maximum masses of that order.
However, when $f_{DM}>6\%$, masses of approximately $2M_{\odot}$ or higher become unstable and the evidence decreases significantly.
For DM with $m_{DM}=100$~MeV or $1$~MeV, the Bayes factor decreases with the fraction due to the disappearance of stable solutions, while for a mass of $10$~MeV the Bayes factor increases at the highest fractions ($9\%$ and $10\%$) because they give rise to new stable configurations.
Finally, for the largest masses ($m_{DM} = 10$~GeV or $m_{DM} = 100$~GeV) the Bayes factor keeps a value very close to $1$.

So, for HESS J1731-347 and PSR J1231-1411, the most favored DM fraction, for a DM mass of $1$~GeV, is the maximum fraction considered in this work (10\%). However, for the XTE J1814-338 CO, the most favored fraction is $6\%$ for the same DM mass.

After marginalizing over the DM fractions, we can determine which value of $m_{DM}$ most strongly favors the DANS hypothesis over the PBNS hypothesis.
In Table~\ref{Tab:Bayes_factors_mDM_separated_analysis} we show the Bayes factors obtained for the three compact objects separately.
The results shown there favor a DM mass of the order of $1$~GeV for the three compact objects separately.

The joint analysis of the three compact objects also shows that a DM mass of $1$~GeV provides the scenario with highest evidence over the pure baryonic one. In Table~\ref{Tab:Bayes_factors_mDM_joint_analysis} we show the results for the logarithm of the joint Bayes factor, which gives the largest value for $m_{DM} = 1$~GeV. 
This result is also reflected in Table~\ref{Tab:mDM_posterior_probabilities}, where we show the evidence for the individual COs and the joint analysis, together with the posterior probability for each mass.

\begin{table*}[ht!]
    \centering
    \begin{tabular}{cccccc}
        \hline\hline
        $m_{DM}$ & $Z^{DANS}_{HESS}$ & $Z^{DANS}_{PSR}$
        & $Z^{DANS}_{XTE}$ & $Z^{DANS}_{joint}$
        & $p(m_{DM}|d_{HESS},d_{PSR},d_{XTE})$ \\
        \hline
        1~MeV      & 0 & $1.56\times 10^{-14}$ & $4.43\times 10^{-22}$ & 0 & 0 \\
        10~MeV     & $1.66\times 10^{-3}$ & $1.99\times 10^{-3}$ & $5.42\times 10^{-32}$ & $3.83\times 10^{-36}$ & $3.23\times 10^{-15}$ \\
        100~MeV    & $1.20\times 10^{-5}$ & $1.94\times 10^{-5}$ & $6.05\times 10^{-21}$ & $3.82\times 10^{-37}$ & $4.06\times 10^{-16}$ \\
        1~GeV   & 0.013 & 0.011 & $9.30\times10^{-22}$ & $1.21\times10^{-22}$ & 0.988 \\
        10~GeV  & 0.004 & 0.004 & $3.82\times10^{-24}$ & $6.75\times10^{-25}$ & 0.006 \\
        100~GeV & 0.004 & 0.004 & $3.82\times10^{-24}$ & $6.75\times10^{-25}$ & 0.006 \\
        \hline
    \end{tabular}
    \caption{Evidence and posterior probabilities for the dark matter masses considered in the joint analysis. $d_{HESS}$, $d_{PSR}$, and $d_{XTE}$ denote the data for the three COs.}
    \label{Tab:mDM_posterior_probabilities}
\end{table*}

\section{\textbf{Dark matter content in a neutron star}}\label{Sec:DM_content}

\begin{figure}[h!]
    \centering
    \includegraphics[width=0.8\linewidth]{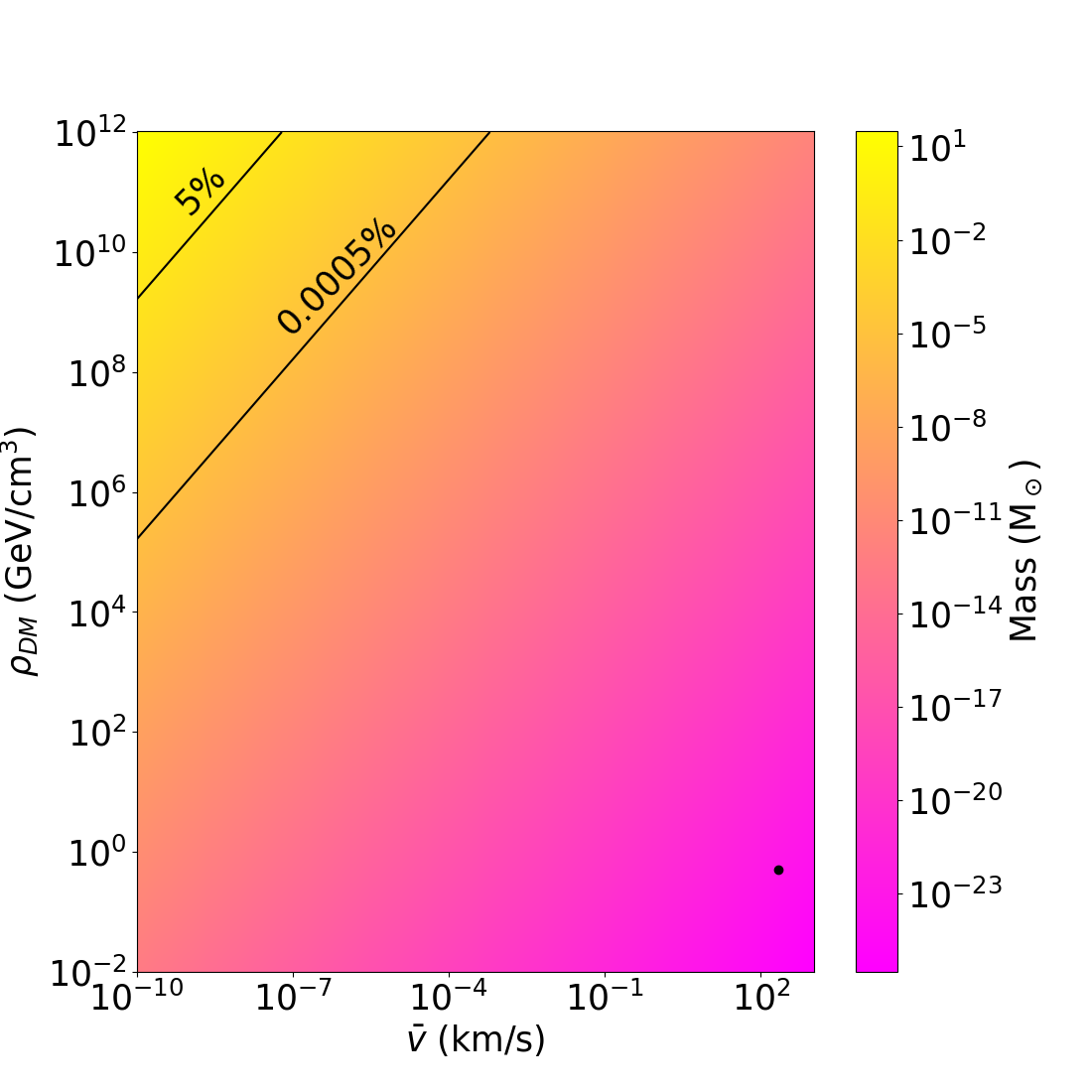}
    \caption{DM captured by an NS for different combinations of ambient density, velocity. The black dot corresponds to average values of the velocity distribution and ambient DM density in the Milky Way.}
    \label{fig:captura}
\end{figure}

Once we have calculated the minimum amount of DM needed to explain the CO measurements for each DM model, it is necessary to determine whether such an amount can be captured during the evolution of an NS.
In order to study the possible DM fractions inside these compact objects, we follow Ref.~\cite{Tinyakov:2021lnt}. 
When a particle from the Galaxy's DM halo scatters with the star, it can lose enough kinetic energy to become gravitationally bound. 
The times that a DM particle crosses the star per unit of time gives an upper bound of the capture rate $\sigma_{cr}$, which does not depend on the energy loss mechanism. 
For an NS in a typical galactic environment, the amount of DM captured is estimated to be approximately $10^{-14}~M_\odot$.
However, higher environment DM densities or smaller velocities can enhance this amount by a few orders of magnitude. 
Nonetheless, there are two factors that often suppress the capture rate. Firstly, not every particle crossing the star gets scattered. The probability of a DM particle scattering with nucleons is governed by the DM-nucleon cross section $\sigma_{\chi N}$. Second, even if the particle gets scattered, if its velocity is large, the energy lost in the collision might not be sufficient for the DM particle to become gravitationally bound. With both factors in mind, the capture rate during the lifetime of the star is:
\begin{equation}
\label{eq:capture rate}
    mF \sim 2.5\times10^{26}~\text{GeV/s}\left(\frac{\rho_{DM}}{0.5~\text{GeV/cm$^3$}}\right)\left(\frac{220~\text{km/s}}{\bar{v}}\right)f
\end{equation}
where $\bar{v}$ is the velocity dispersion of the DM particles, $\rho_{DM}$ is the ambient DM density, and $f$ is the fraction of particles that scatter among all particles crossing the star:
\begin{align}
    &f = \sigma_{\chi N}/\sigma_{cr} \quad \text{if }\quad  \sigma_{\chi N}<\sigma_{cr} \\
    &f = 1 \quad \text{otherwise}
\end{align}
For weak interactions ($f=10^{-10}$) and an NS with a lifetime of 10~Gyr in the Milky Way, the estimated DM mass captured during the lifetime of the star is of the order of $10^{-23}~M_\odot$.
In Fig.~\ref{fig:captura} we show different combinations of ambient DM density and DM-particle velocity, together with the resulting captured mass, for two different NS ages and interaction strengths.
It is shown that, with conventional capture methods, for the NS to have $0.0005\%$ of its content as DM, the ratio between the velocity distribution and the DM ambient density should be
\begin{equation}
    10^{10} \text{ years, } f=10^{-10}: \quad \left. \left(\frac{\rho_{DM}}{\bar{v}}\right)=10^{17} \left(\frac{\rho_{DM}}{\bar{v}}\right)\right|_\text{average} 
\end{equation}

There are other scenarios that can enhance the DM content inside the CO.
One of them is the DM capture of the progenitor star before the CO formation.
However, even though the progenitor can accumulate a DM mass of $\sim 10^{-11}$~$M_{\odot}$ in this case \cite{Tinyakov:2021lnt}, there is no evidence that this DM content remains bound to the resulting NS after the supernova explosion. Therefore, to be conservative regarding the amount of DM present in an NS, we consider only the DM captured during the NS's lifetime.

On the other hand, fractions of approximately $1\%$ have been achieved in models with asymmetric dark matter \cite{Parmar:2025bux} or DM conversion inside the NS \cite{Bastero-Gil:2024kjo}, but these scenarios remain uncertain.

\section{SUMMARY AND CONCLUSIONS}\label{Sec:Summary}

In this work, we have studied the possibility that three COs with anomalous masses and radii are DANSs.
For the DM model, we use a relativistic free Fermi gas with different DM masses. 
Using the two-fluid formalism, we studied the mass-radius diagram and stability of a DANS for a fixed DM fraction inside the baryonic radius.
In our analysis, we found two different DANS configurations. 
The first one is when the DM component stays inside the baryonic radius (core configuration), and emerges when $m_{DM} \geq 1$~GeV. 
The second one is the halo configuration, where the DM component expands outside the baryonic radius, and appears when $m_{DM} \leq 100$~MeV.

The latter produces extended haloes that give rise to unstable configurations in the mass region of the anomalous COs, except for some DM fractions and baryonic EoSs. Additionally, these massive DM haloes would modify the measured masses of the binary systems to which PSR J1231-1411 and XTE J1814-338 belong.

We performed a Bayesian analysis to compare the DANS and PBNS hypotheses, considering different values of the DM mass.
We reconstructed the likelihoods needed in the analysis from the published posteriors for the three COs and their reported priors.
We find that the DANS hypothesis is favored over the PBNS hypothesis for $m_{DM}=1$~GeV, since any DM content inside the baryonic object shifts the radius of the M-R curve toward regions of higher probability density while preserving the stability of the NS.
Our work shows that the Bayesian analysis yields a DM mass of $1$~GeV as the most probable among the cases studied under the DANS hypothesis.

In order to produce significant modifications of the M-R curve, the compact object needs to contain a DM fraction of the order of 1\%. 
We investigated the possibility of a DANS existing with the required fractions for the different masses considered in this work.
In this regard, we show that the required DM content is much higher than the mass captured during the typical lifetime of an NS in a typical galactic environment.
To be able to reach the requiered fractions, the COs would need to be in environments with exceptionally large ambient DM density or slow DM particles. 
Even DM capture by the progenitor star could not explain such a large amount of DM in standard galactic environments.

Other works suggest that compact objects may contain a high DM fraction for different reasons. One possible scenario that could explain DM fractions of the order of $(1-10)\%$ is the presence of DM clumps that could accrete baryonic matter due to free streaming \cite{Tolos:2015qra}, or a dark-star companion \cite{Eby:2015hsq,Maselli:2017vfi,PhysRevD.92.063526} from which the compact object could accrete sufficient DM.

Other scenarios can enhance the DM content inside the CO, such as models with asymmetric dark matter \cite{Parmar:2025bux} or DM conversion inside the NS \cite{Bastero-Gil:2024kjo}, but their viability remains uncertain.
Therefore, although the DANS hypothesis gets favored over the PBNS hypothesis in the Bayesian analysis for some cases, the interpretation of the COs as DANS is in tension with standard capture scenarios by several orders of magnitude.
Although DANSs in typical galactic environments have DM fractions too low to produce visible effects in the mass-radius diagram, other observables may allow us to probe the DM content of the star. For example, the tidal deformability is expected to decrease in the presence of a dark matter core \cite{Ellis:2018bkr,Sagun:2021oml} and increase when the DM forms a halo \cite{Nelson:2018xtr}.
Future work could also investigate the effects of a massive, extended dark matter halo on X-ray measurements.

\newpage
\begin{acknowledgments}
    We thank Laura Tolos and Marina Cermeño for their insightful comments and discussions. We also thank Yves Kini for sharing the posterior of XTE J1814-338, Valery Suleimanov and Victor Doroshenko for their assistance in reproducing the posterior of HESS J1731-347, and Mingyu Ge for helping us find the data needed to reconstruct the posterior of PSR J1231-1411.
We acknowledge partial financial support to the Grant PID2022-136510NB-C32 funded by
MCIN/AEI/10.13039/501100011033/ and FEDER, UE,
and the project EPU-INVUAH/
2023/003 funded by the University of Alcal\'a.
 The codes and equations of state used in this work are available upon request.
\end{acknowledgments}

\appendix
\section{Mass--radius diagrams for all equations of state}
\begin{figure}[h!]
  \centering
      \includegraphics[width=0.45\textwidth]{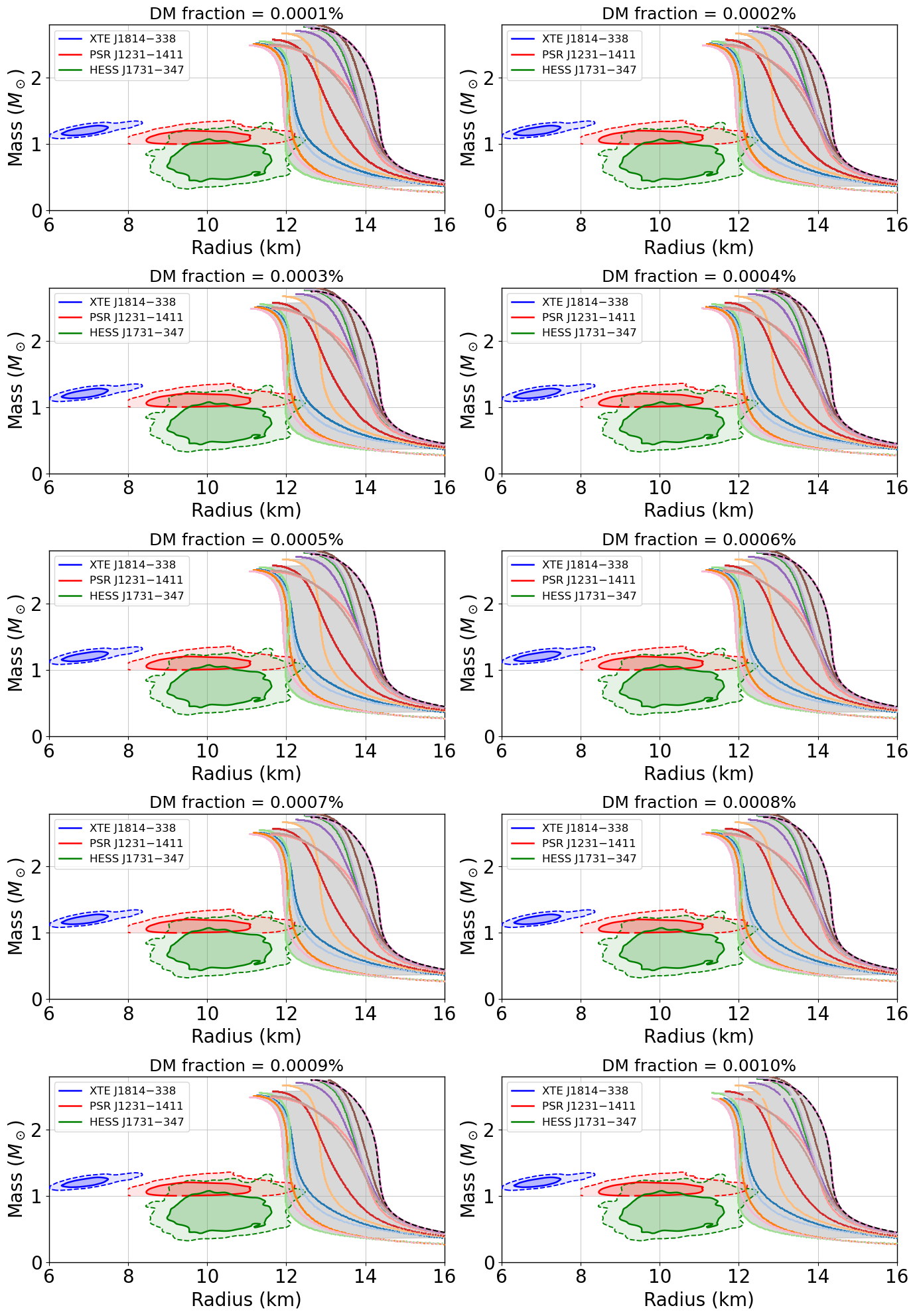}\hspace{1cm}  
      \includegraphics[width=0.45\textwidth]{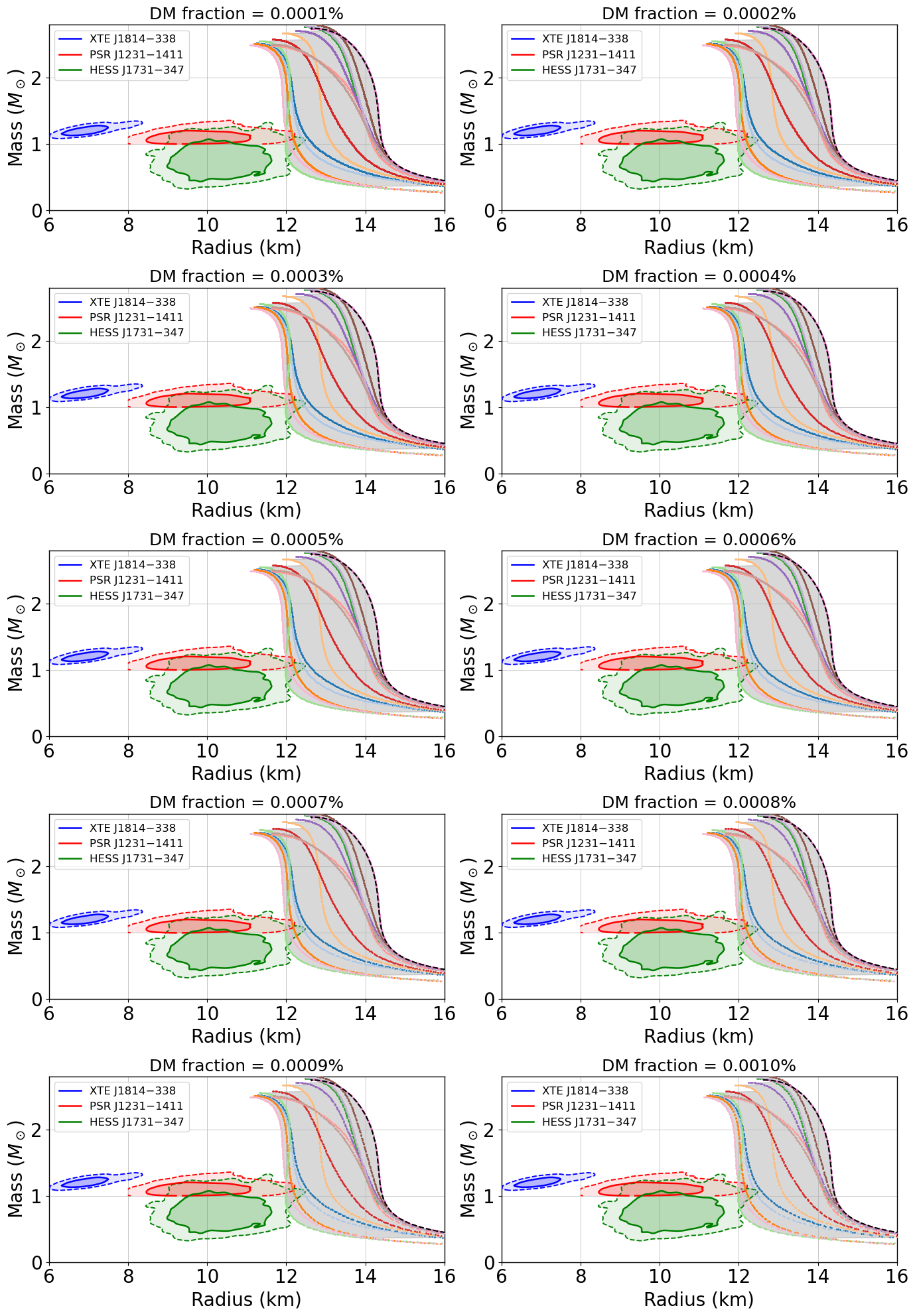} 
  \caption{Stable configurations for DM fractions in the range of $(10^{-3}-10^{-4})\%$ with a DM mass of $100$~GeV (left panel) and $10$~GeV (right panel). The black dashed curve corresponds to the rightmost curve of Fig.~\ref{Fig:Baryonic} (no DM content) and is shown for comparison. The green, red, and blue contours correspond to the HESS J1731-347, PSR J1231-1411, and XTE J1814-338 posteriors at the 68\% (solid) and 95\% (dashed) credible levels.}
  \label{Fig:100GeV}
\end{figure}

\begin{figure}[h!]
  \centering
      \includegraphics[width=0.45\textwidth]{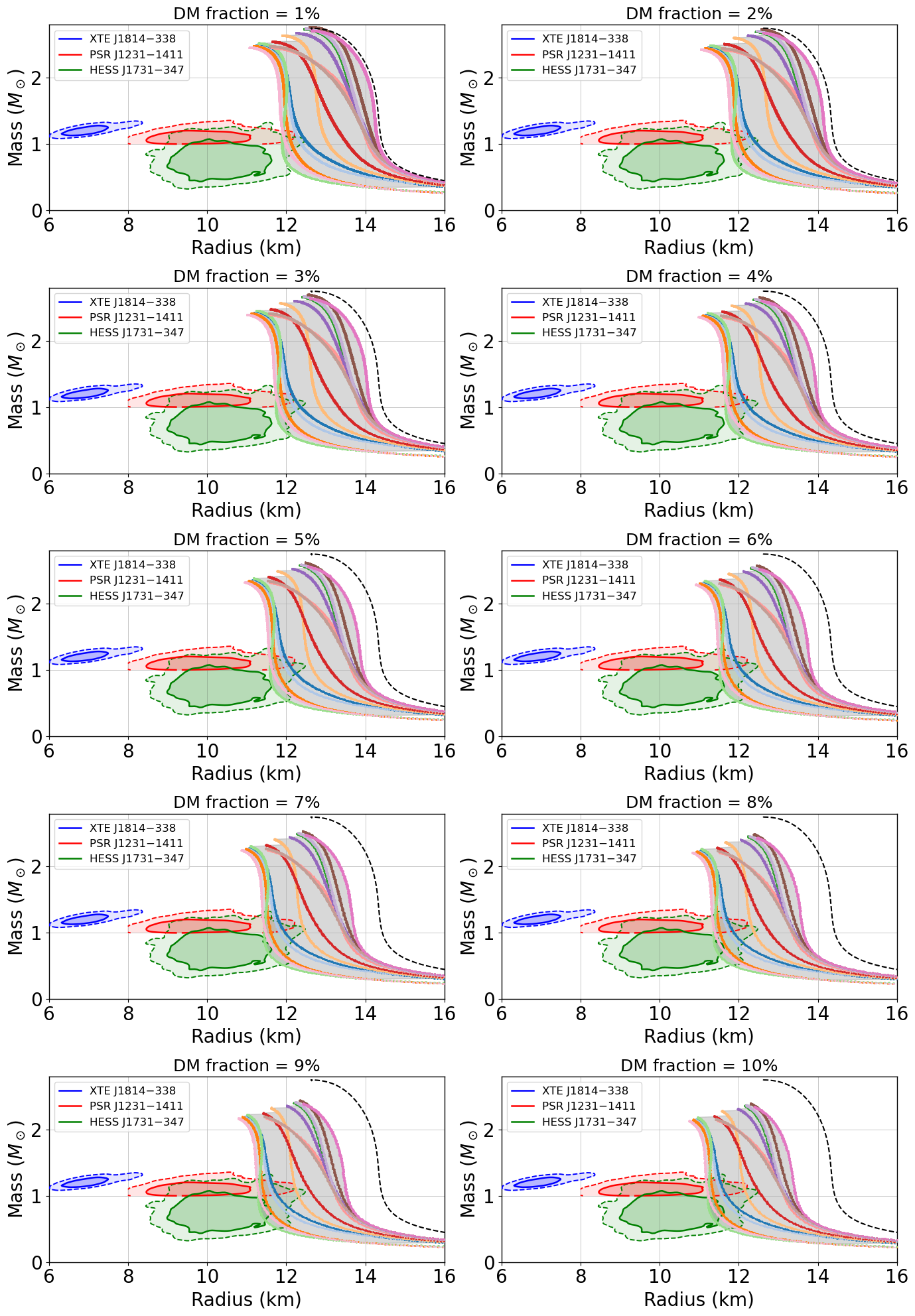}\hspace{1cm}  
      \includegraphics[width=0.45\textwidth]{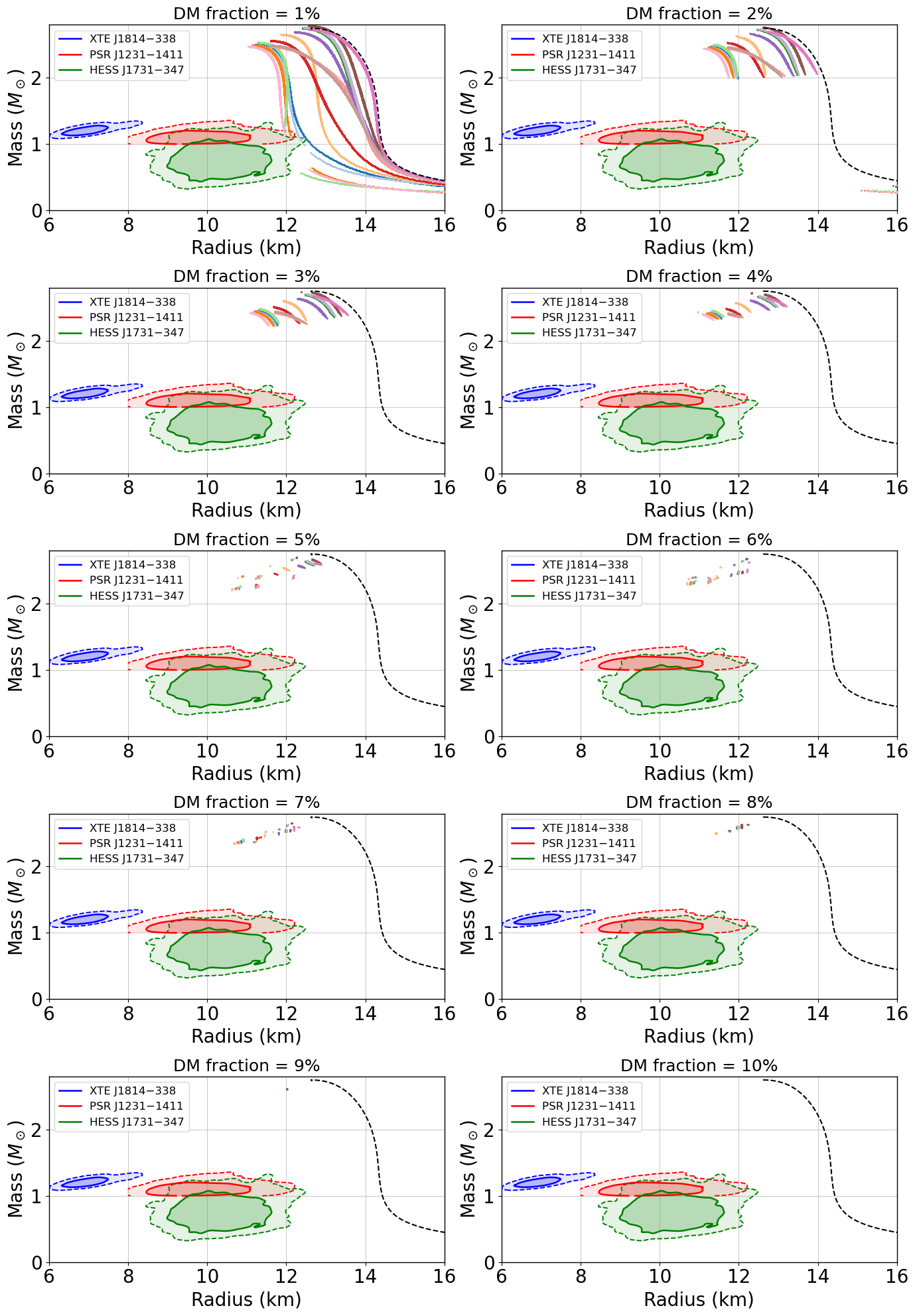} 
  \caption{Stable configurations for DM fractions in the range of $(1-10)\%$ with a DM mass of $1$~GeV (left panel) and $100$~MeV (right panel). The black dashed curve corresponds to the rightmost curve of Fig.~\ref{Fig:Baryonic} (no DM content) and is shown for comparison. The green, red, and blue contours correspond to the HESS J1731-347, PSR J1231-1411, and XTE J1814-338 posteriors at the 68\% (solid) and 95\% (dashed) credible levels.}
  \label{Fig:1GeV}
\end{figure}

\begin{figure}[h!]
  \centering
      \includegraphics[width=0.45\textwidth]{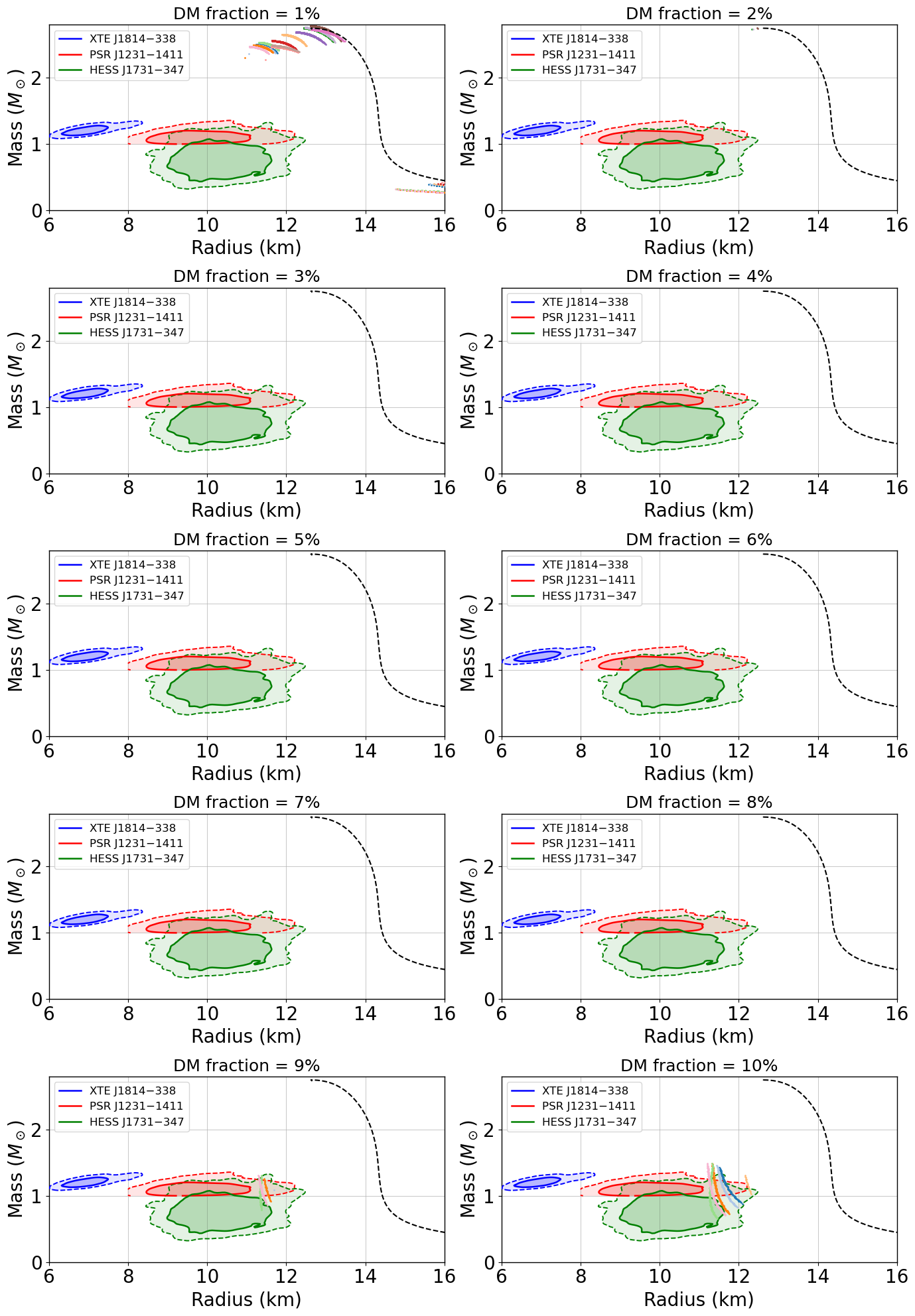}\hspace{1cm}  
      \includegraphics[width=0.45\textwidth]{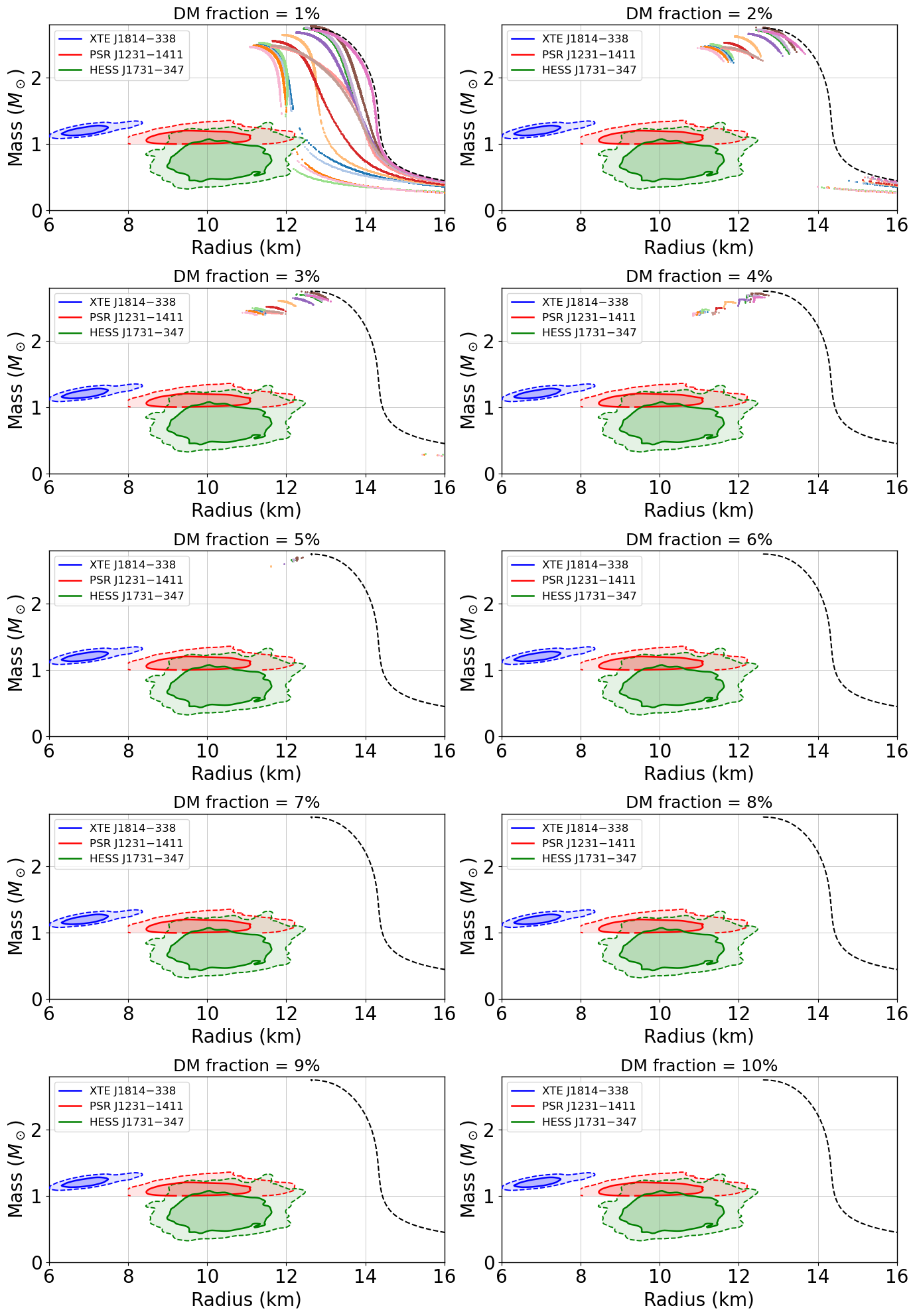} 
  \caption{Stable configurations for DM fractions in the range of $(1-10)\%$ with a DM mass of $10$~MeV (left panel) and $1$~MeV (right panel). The black dashed curve corresponds to the rightmost curve of Fig.~\ref{Fig:Baryonic} (no DM content) and is shown for comparison. The green, red, and blue contours correspond to the HESS J1731-347, PSR J1231-1411, and XTE J1814-338 posteriors at the 68\% (solid) and 95\% (dashed) credible levels.}
  \label{Fig:10MeV}
\end{figure}

In Figs. \ref{Fig:100GeV}-\ref{Fig:10MeV}, we present the stable sequences obtained for different DM fractions and particle masses across all the baryonic EoSs considered in this work. This comparison allows us to 
asses the systematic dependence of our results on the choice of baryonic EoS.
\label{App:M-R_diagrams_all_EoS}

\clearpage
\bibliography{biblio}
\bibliographystyle{aasjournalv7}

\end{document}